\documentclass[preprint2]{emulateapj}

\usepackage{graphicx}
\usepackage{natbib}
\usepackage{color}
\usepackage{array}
\usepackage{rotating}
\usepackage{amsmath}
\usepackage{amssymb}
\usepackage{lineno}
\usepackage{appendix}
\usepackage{adjustbox}

\newif\ifarxiv

\arxivtrue

\ifarxiv
\else
\linenumbers
\fi

\usepackage[version=3]{mhchem} 

\def\E{\mathrm{e}}

\definecolor{blueish}{rgb}{0.2, 0.8, 0.8}

\shorttitle{Redox evolution of low mass planets}
\shortauthors{Wordsworth}

\begin{document}


\title{Redox evolution via gravitational differentiation on low mass planets: implications for {abiotic oxygen}, water loss and habitability}


\author{R.~D.~Wordsworth}
\affil{School of Engineering and Applied Sciences, Harvard, Cambridge, MA 02138, USA}
\affil{Department of Earth and Planetary Sciences, Harvard, Cambridge, MA 02138, USA}
\email{rwordsworth@seas.harvard.edu}

\author{L.~K.~Schaefer}
\affil{School of Earth and Space Exploration, Arizona State University, Tempe, AZ 85287, USA}

\author{R.~A.~Fischer}
\affil{Department of Earth and Planetary Sciences, Harvard, Cambridge, MA 02138, USA}



\begin{abstract}
The oxidation of rocky planet surfaces and atmospheres, which arises from the twin forces of stellar nucleosynthesis and gravitational differentiation, is a universal process of key importance to habitability and exoplanet biosignature detection. Here we {take a} generalized approach to this phenomenon. Using a single parameter to describe redox state, we model the evolution of terrestrial planets around nearby M-stars and the Sun. Our model includes atmospheric photochemistry, diffusion and escape, line-by-line climate calculations and interior thermodynamics and chemistry. In most cases we find abiotic atmospheric \ce{O2} buildup {around M-stars during the pre-main sequence phase} to be much less than calculated previously, because the {planet's magma ocean} absorbs most oxygen liberated from \ce{H2O} photolysis. However, loss of {non-condensing} atmospheric {gases after the mantle solidifies} remains a significant {potential} route to abiotic {atmospheric} \ce{O2} subsequently. In all cases, {we predict that} exoplanets that receive lower stellar fluxes, such as LHS1140b and TRAPPIST-1f and g, have the lowest probability of abiotic \ce{O2} buildup and hence may be the most interesting targets for future searches for biogenic \ce{O2}. Key remaining uncertainties can be {minimized} in future {by comparing our predictions} for the atmospheres of hot, sterile exoplanets such as GJ1132b and TRAPPIST-1b and --c {with observations}. 
\end{abstract}

\maketitle

\section{Introduction}

Following the recent discoveries of nearby exoplanets with masses  in the 1-10~$M_E$ range, we are faced with the exciting prospect that in the near future, characterization of the atmospheres of rocky planets outside the solar system will be possible \citep{Udry2007,BertaThompson2015,Gillon2016,Anglada2016,Gillon2017,Dittmann2017}. Some of these planets, such as GJ1132b or TRAPPIST-1b and c, receive a greater stellar flux than Venus and hence are likely to have hot, possibly molten surfaces \citep{BertaThompson2015,Gillon2016,Schaefer2016}. Others, such as Proxima Centauri b and LHS1140b, are potentially habitable to Earth-like life, depending on their atmospheric composition \citep{Kasting1993,Wordsworth2010b,Pierrehumbert2011b,Kopparapu2013,Barnes2016,Turbet2016}.

Development of a general framework for predicting the atmospheric composition of rocky planets is one of the major theoretical challenges of the field in the coming years. For high-mass planets, atmospheres are invariably hydrogen-dominated, and composition at a given pressure is dominated by a balance between thermo- and photochemical effects \citep[e..g,][]{Moses2011}. For low mass planets, the bulk atmospheric composition is considerably harder to predict, because the external boundary conditions (escape to space, delivery from planetary embryos and comets, and outgassing / subduction) have a fundamental and still poorly constrained influence \citep[e.g,][]{Morbidelli2000,HirschmannWithers2008,Lammer2008,Lenardic2012,Wordsworth2013b,Dong2017}. 

Given the complexity of the problem, simplifying assumptions are essential for progress. One useful approach is to limit the number of chemical elements in a model to the bare minimum needed to capture essential features. For example, galactic elemental abundances are such that among the non-noble volatiles, H, C, N, O and S can be expected to dominate the composition of almost any planetary atmosphere receiving a stellar flux within an order of magnitude of that received by Earth. However, even for atmospheres restricted to just these elements, the phase space of composition remains extremely large, as evidenced by the diversity of atmospheres in our own solar system. 

One potentially fruitful approach is to characterize every atmosphere in terms of a single chemical variable. Appropriately defined, atmospheric \emph{redox state} is particularly useful, because of the dominant controlling role of redox in atmospheric and surface chemistry \cite[e.g.,][]{Yung1999}. Redox evolution is also extremely important to astrobiology. First, formation of prebiotic molecules, and hence biogenesis, proceeds most readily on planets with weakly or highly reducing atmospheres and surfaces  \citep{Miller1959,Zahnle1986,Powner2009,Tian2011,Ranjan2017}. Second, the highly oxidized state of Earth's present-day atmosphere and much of its surface is a product of the biosphere, and hence \ce{O2} has potential as a \emph{biosignature}, or unique sign of life \citep{Selsis2002,Kaltenegger2010,Seager2012,Zahnle2013}. Nonetheless, it has recently been shown that abiotic processes may lead to buildup of \ce{O2}-dominated atmospheres on planets that lack life in some cases \citep{Wordsworth2014,Luger2015,Schaefer2016}. These cases constitute `false positives' for life that require careful study to discriminate them from biologically generated atmospheres \citep{Domagal2014,Schwieterman2015,Meadows2016}. A robust understanding of the factors that control a planet's surface and atmospheric redox evolution is therefore critical for future observational searches for life on other worlds.

Here we take a generalized approach to this problem. We focus on abiotic processes that can cause irreversible oxidation of planetary surfaces and atmospheres, because they are most relevant to biosignature definition and to prebiotic chemistry. Extending our previous specific study of the atmospheric evolution of the exoplanet Gliese 1132b \citep{Schaefer2016}, we model both interior-atmosphere exchange and the escape to space of key atomic species.  In Section~\ref{sec:general}, we discuss planetary oxidation from a general perspective. In Section~\ref{sec:escape}, we discuss atmospheric escape, and in Sections~\ref{sec:melty} and \ref{sec:crusty} we discuss coupling between the atmosphere and planetary interior. In Sections~\ref{sec:trap}-\ref{sec:other_spec} we discuss the important issue of \ce{H2O} cold-trapping and the role of hydrogen-bearing species other than \ce{H2} and \ce{H2O}. The key findings of this work and future directions are discussed in Section~\ref{sec:diss}-\ref{sec:future}.

\section{A generalized approach to planetary oxidation}\label{sec:general}

\begin{figure}[h]
	\begin{center}
\ifarxiv
		{\includegraphics[width=3.5in]{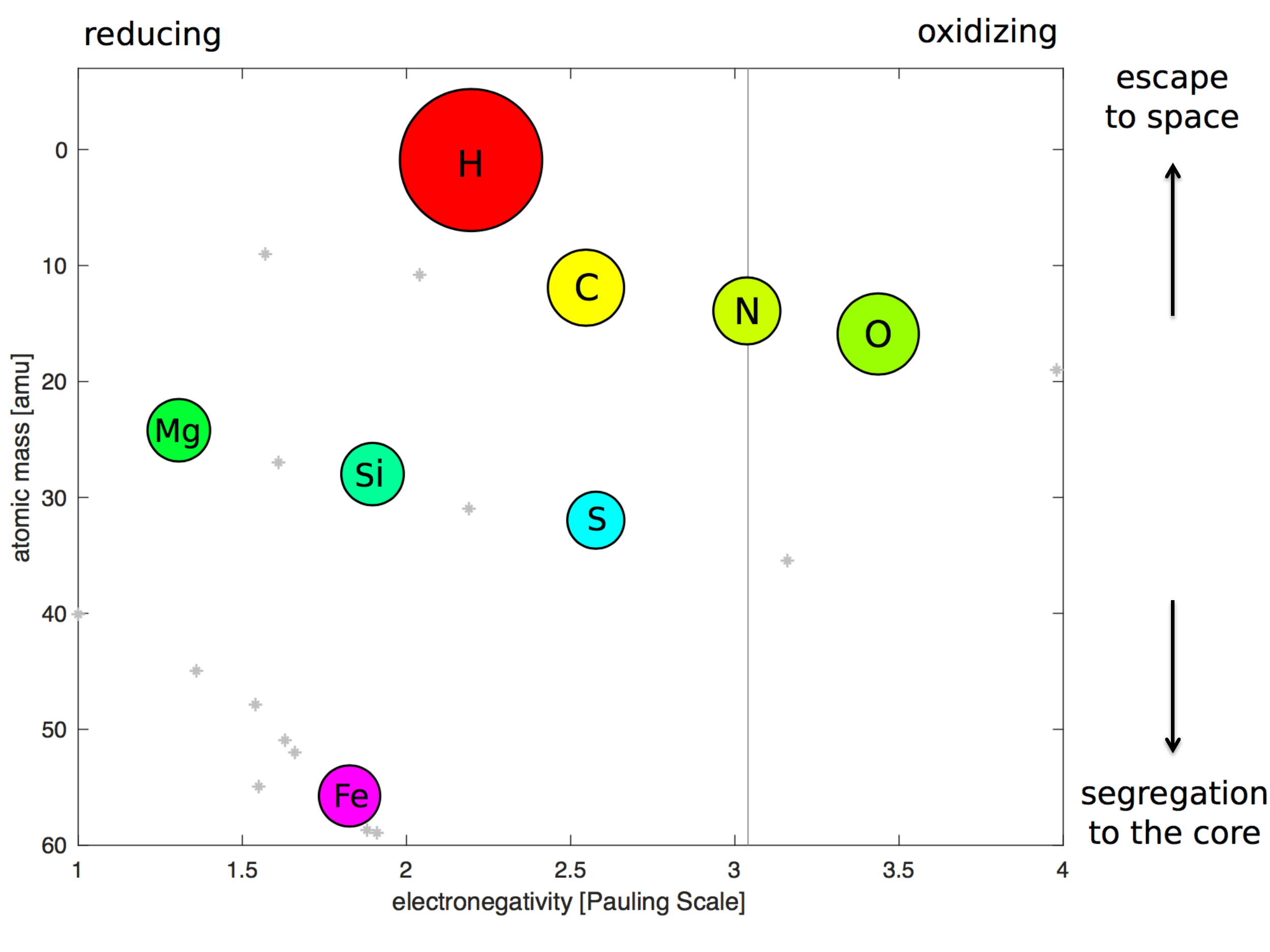}}
\else
		{\includegraphics[width=3.5in]{figures/redox_schematic_5.pdf}}
\fi
	\end{center}
	\caption{ Plot of  atomic mass vs. electronegativity for the major elements in the solar system, with the size of the circle corresponding to elemental abundance. Gray asterisks denote elements with abundances lower than 10\% that of Si. Escape to space is dominated by lower mass elements (particularly H), while higher mass elements (particularly Fe) tend to segregate to the planet's core. The intermediate mass, more electronegative elements C, N and O dominate the atmospheres of the rocky solar system planets.  Abundance data is from \cite{Lodders2003} and electronegativity data is from \cite{Pauling1967}.}
\label{fig:schematic_2}
\end{figure}

While the idea that rocky planets can oxidize abiotically via \ce{H2O} photolysis followed by hydrogen loss to space is well-developed \cite[e.g., ][]{Oparin1938,Kasting1983,Chassefiere1996,Wordsworth2014,Luger2015,Schaefer2016}, the simplicity of the physics driving oxidation is often obscured by the host of other complex effects that can sculpt planetary atmospheres, not to mention the interplay between redox and life on Earth. To understand why surface oxidation should be expected as a general rule, it is useful to compare the reducing power and atomic masses of the major elements that make up low mass planets. Figure~\ref{fig:schematic_2} shows a plot of the major solar system elements as a function of their electronegativity according to the Pauling scale \citep{Pauling1967} and their atomic mass.  The size of each circle scales with the logarithm of the element's abundance \citep{Lodders2003}. Only elements with solar system abundances of $4\times 10^{-6}$ or more relative to hydrogen are displayed.

If we treat solar system element abundances as a proxy for those in exoplanet systems, primordial planetary atmospheres are likely to be dominated by one reducing (low electronegativity), highly volatile element (H), one oxidizing (high electronegativity) element of intermediate mass (O) and two intermediate elements that are  less abundant (C and N). The heaviest major element, Fe, has significant reducing power. The intermediate mass elements (Na to Ca) have generally low electronegativity but tend to combine rapidly with the more abundant oxygen on condensation in the protoplanetary disk, and subsequently remain in the crust and mantle for all but the hottest planets. For most compounds there is a strong correlation between mean molecular mass and density at a given pressure, and so  iron preferentially accumulates in the core and hydrogen preferentially escapes to space. 
On terrestrial-type planets, gravitational segregation therefore \emph{always} acts to drive reducing power away from the surface and atmosphere. 

Simple as this principle is, it is interesting to note that it depends entirely on the selective effects of stellar nucleosynthesis. The abundance of carbon and oxygen relative to elements such as lithium, beryllium and boron is a consequence of the physics of helium burning in late-stage stars \citep{Clayton1968}. In a hypothetical universe where lithium was a dominant product of stellar fusion, hydrogen loss would cause an \emph{increase} in the total reducing power of a planet's surface. Planetary oxidation is probably crucial to the origin and development of complex life, so the fact that lithium is not a major element is a rather fascinating and fortunate outcome of nuclear physics.

If segregation of iron to a planet's core was perfectly efficient and escape of hydrogen to space was independent of atmospheric composition, constructing a general theory of planetary redox evolution would be easy. However, the escape of hydrogen is strongly dependent on its abundance and chemical form in the atmosphere, the mantle iron content in rocky planets is significant, and the rate of transport of oxygen into the planetary interior is a strong function of the mantle thermal state. In the following sections, we describe our approach to modeling each of these processes.

\subsection{A single variable for redox state}

For convenience, we begin by defining a single redox variable. We first place all elements on an electronegativity scale and set the zero point equal to the electronegativity of nitrogen\footnote{This is a somewhat arbitrary choice, but it fits our emphasis on the interaction between the abundant oxidizing element O and the other key constituents. It also fits with the fact that nitrogen is not a major reducing or oxidizing agent compared to H, O or Fe.}. We then categorize each element according to the maximum number of electrons it will exchange in interaction with an element on the other side of the electronegativity divide\footnote{Emphasis on the most abundant elements here allows us to ignore the wider range of oxidation states that may occur in combination with other elements, e.g. \ce{Fe^{6+}} in \ce{K2FeO4}. 
These states are important to chemistry in general but not to bulk planetary evolution.}. 
For any planetary reservoir, the total oxidizing power $N$ can then be calculated as 
\begin{equation}
N = \sum_iN_i p_i
\end{equation}
where $N_i$ is the number of atoms of a given element and $p_i$ is the element's oxidizing potential. Frequently, we will be working with large numbers of atoms, so it is convenient to express $N$ in terms of the total amount of accessible electrons in the hydrogen in Earth's oceans ($N_{e,\mbox{TO}} = 9.15\times10^{46}$). 
The oxidizing potential for ten major elements, alongside their solar and bulk silicate Earth (BSE) abundances, are given in Table~\ref{tab:redox_defs}. This approach bears some similarity to schemes proposed to describe the redox budget of planetary atmospheres in the past \citep[e.g., ][]{Kasting1998}, but its direct link to elemental electronegativity allows for more systematic classification.

\begin{table}[h]
\centering
\begin{tabular}{cccc}
\hline
\hline
Element &  $p_i$ & Solar abundance & BSE abundance  \\
\hline
H & {--1} & 24300 & $< 4.0\times10^{-2}$  \\
C & --4 & 7.08 & $5.9\times10^{-3}$  \\
N & 0& 1.95 & $1.6\times10^{-5}$  \\
O  & +2& 14.13 & $3.76$ \\
Mg  & --2& 1.02 & $1.28$ \\
Al  & {--3}& 0.084 & $0.10$ \\
Si  & --4& 1.0 & 1.0 \\
S  & --{6}& 0.45 & $<1.0\times10^{-3}$ \\
Ca  & {--2}& 0.063 & $0.075$ \\
Fe  & {--3}& 0.84 & $0.14$ \\
\hline
\hline
\end{tabular}
\caption{Oxidizing potential (as defined in the main text) and solar and bulk silicate Earth  (BSE) abundances (atomic fraction) of the major planetary elements. Abundances are defined relative to Si and are from \cite{Lodders2003} (solar), \cite{Marty2012} (BSE H, C and N) and \cite{Javoy1999} (BSE, other elements).}
\label{tab:redox_defs}
\end{table}

\begin{figure}[h]
	\begin{center}
\ifarxiv
		{\includegraphics[width=2.5in]{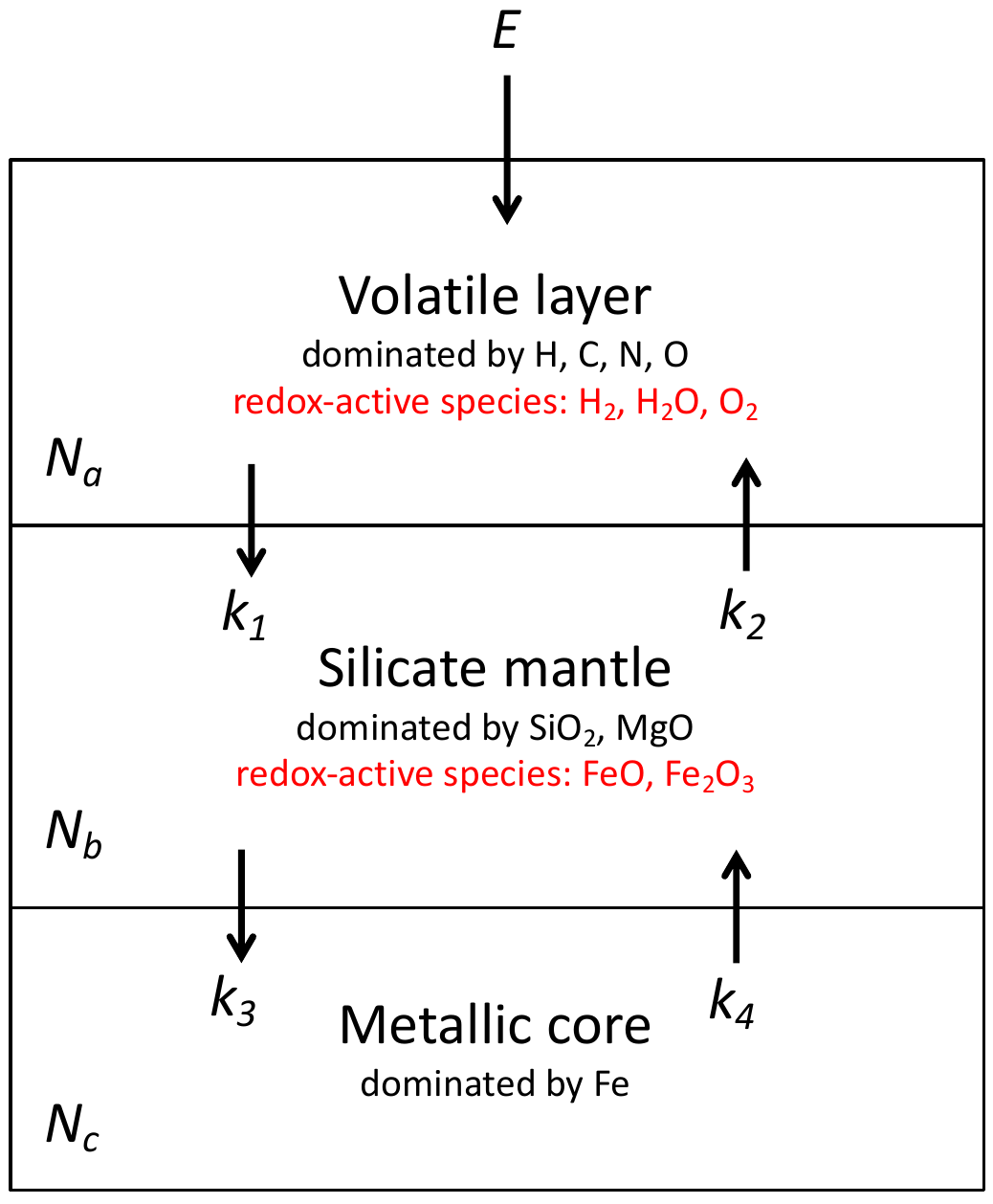}}
\else
		{\includegraphics[width=2.5in]{figures/schematic_4.pdf}}
\fi
	\end{center}
	\caption{Schematic of the box model approach to planetary redox flow, with layers  defined in terms of bulk composition. Differential escape preferentially removes hydrogen and hence represents a positive flux of net oxidizing power.}
\label{fig:box_model_schematic}
\end{figure}

After formation, all planets in the 1-10 Earth mass range are predicted to differentiate into an iron-dominated core, a silicate mantle, and a volatile layer containing lighter species. In the simplest terms, the abiotic redox evolution problem can then be framed as the exchange of oxidizing power between these three reservoirs (Figure~\ref{fig:box_model_schematic}), such that
\begin{eqnarray}
\dot N_{a} &=&  - k_1 N_{a} + k_2 N_{b} + E(t) \label{eq:basic1}\\
\dot N_{b} &=& + k_1 N_{a} - (k_2+k_3) N_{b} + k_4 N_{c}  \label{eq:basic2}\\
\dot N_{c} &=& + k_3 N_{b}- k_4 N_{c}   .\label{eq:basic3}
\end{eqnarray}
Here $N_a$, $N_b$ and $N_c$ are the total oxidizing power of the volatile layer, silicate mantle and core, the $k_{1-4}$  are exchange terms {(see Figure~\ref{fig:box_model_schematic})}, and $E(t)$ captures oxidation due to preferential atmospheric escape of hydrogen. Situations where the value of $N_a$ become positive are of particular importance to us, because this is when \ce{O2} and other oxidizing species will begin to accumulate in the volatile layer.

The size of the exchange terms $k_{1-4}$ is strongly dependent on the phases of the layers in question. For example, on present-day Earth with a solid silicate mantle, the value of $k_1$ and $k_2$ is of order Gy$^{-1}$. In contrast, on a newly formed planet with a liquid silicate layer (magma ocean), $k_1$ and $k_2$ have characteristic values of weeks$^{-1}$ to days$^{-1}$ \citep{Solomatov2007}. However, because magma oceans may only extend across a given region in the mantle \citep{Abe1997,Lebrun2013,Schaefer2016}, treatment of separate solid and liquid silicate reservoirs is important.

If magma oceans solidify from deep in the mantle upwards, exchange rates between the core and mantle ($k_{3-4}$) will be low as soon as the main period of planetary differentiation is complete, and \eqref{eq:basic3} can be neglected. Here we make the simplification that the core formation period occurs quickly and hence sets the initial condition for the mantle oxidizing power $N_b$ in our model. After this point, we assume zero exchange between regions $b$ and $c$. This assumption is most justified for planets around M-stars, which are expected to have long-lived magma oceans due to their host stars' extended pre-main sequence phases.  The physics and chemistry of core formation in general is discussed next.

\subsection{The lower boundary condition: Core formation and initial mantle composition}

During core formation, chemical interactions between molten mantle and core materials at high pressures ($p$) and temperatures ($T$) set the redox state of the mantle, and thus $N_b$. As a planet grows larger, the {average pressure and temperature} of metal-silicate equilibration, which likely occurs in or at the base of the magma ocean in the silicate layer \citep{Rubie2003,Stevenson1981}, are both generally considered to increase \citep{Fischer2017,Rubie2011}. 

At very high pressures and temperatures, some elements that are less electronegative than Fe on the Pauling scale will accept electrons (give up oxygen atoms), becoming neutral and dissolving into the core as metals. This complicates the simple picture of clean separation between elements suggested by Figure~\ref{fig:schematic_2}. The most important of these less electronegative elements in terms of planetary redox changes is silicon \citep{Fischer2015,Siebert2012,Tsuno2013}, which undergoes the reaction $\ce{Si^{4+} + 4e^- \to Si^0}$. This half of the redox reaction is balanced by more electronegative (Pauling scale) elements donating electrons to oxygen atoms, forming silicates and oxides and entering the mantle. Most importantly, Fe oxidizes from its metallic form to enter the mantle as ferrous iron, leading to the overall reaction
\begin{equation}
\ce{Si^{4+} + 2Fe^0  \to Si^0 + 2Fe^{2+}}.\label{eq:core_form2}
\end{equation}
This transfer of electrons from Fe to Si is the primary mechanism for increasing a planet's mantle FeO inventory during core formation \citep{Fischer2017,Ringwood1959,Rubie2011,Rubie2015}. It can change the composition of the mantle significantly, increasing the FeO content by a factor of around three \citep{Fischer2017} or more \citep{Rubie2011,Rubie2015}. However, reactions occurring during core formation do not significantly alter the composition of the core itself, except for the addition of some light elements like Si and O; its iron content does not change significantly for planets near an Earth mass \citep[e.g.,][]{Fischer2017,Rubie2011,Rubie2015}. 

In planets that are larger than an Earth mass, the pressures and temperatures of metal--silicate equilibration will be higher. At higher pressures and particularly at higher temperatures, reaction (\ref{eq:core_form2}) will proceed farther to the right  \citep{Fischer2015,Siebert2012}, leading to larger redox changes, a higher mantle FeO content, and hence a more negative initial value of $N_b$. Plausibly, planets that are hotter during formation for other reasons (such as more energetic impacts during accretion), will also have higher mantle FeO content. 

{
In core formation studies, it is standard to refer to the addition of FeO to the mantle as a net \emph{oxidation} of the mantle, because Fe loses electrons to O when it is removed from the core. However, from an atmospheric/surface perspective, the most important outcome of reaction (\ref{eq:core_form2}) is that the iron added to the mantle can be further oxidized to \ce{Fe^{3+}}, and hence constitutes a potential sink of oxidizing power (more negative value of $N_b$ in our scheme). This discrepancy of terminology is probably linked to the fact \ce{FeO} is the most oxidized iron species under core mantle boundary conditions, while \ce{Fe2O3} is the most oxidized form of iron on planetary surfaces.}

Existing models of core-mantle equilibration during accretion for the inner solar system planets yield {FeO mantle abundances} ranging from 6-20~wt\%, in approximate agreement with estimates of Earth, Venus, Mars and Mercury's actual mantle iron content \citep{Fischer2017}. Here we vary the initial mantle FeO content from 0 to 20~wt\%. Nonetheless, we regard 5~wt\% as a plausible lower limit in all but the most extreme cases.

Though Si plays an important role in planetary redox during core formation, mantle silicon is subsequently bonded with oxygen and mainly remains in the silicate mantle without any valence changes. Likewise, Mg, Al, and Ca readily bond with the more abundant O in protoplanetary disks and subsequently mainly remain in the silicate mantle. Here these elements as well as Si are neglected in the overall redox budget, but Fe is included. S is relatively scarce in the bulk silicate Earth, so we also neglect it here, although it may play an important role in certain cases. Finally, C and N species (particularly \ce{CO2}, \ce{CH4} and \ce{N2}) can have important atmospheric effects, but their direct contribution of these elements to the redox budget is also typically smaller than that of O and H. The contributions of \ce{CO2} and \ce{N2} to the greenhouse effect and to atmospheric cold-trapping of \ce{H2O} are considered in Sections~\ref{sec:melty}-\ref{sec:crusty}. However, in the redox evolution modeling, the only active species we allow are H, O and Fe.

\subsection{Redox disproportionation of Fe}\label{sec:redox}

Besides direct interaction of the silicate melt with the core, a second potentially important influence {on planetary redox evolution} during the late stages of core formation is $\ce{Fe}$ redox disproportionation. This becomes important when the crystallization pressures are greater than about 24~GPa\footnote{For comparison, Earth's core-mantle boundary pressure is approximately 140~GPa.} and the mineral bridgmanite [\ce{MgSi O3}] is stable \citep[e.g., ][]{Fei2004}. At these pressures, \ce{Fe3+} can incorporate into bridgmanite in the reaction \citep{Frost2004}
\begin{equation}
\ce{3Fe^{2+}  \to Fe^0  + 2Fe^{3+}}\label{eq:redox_disprop}
\end{equation}
This reaction is more likely when abundances of Al are high, due to a coupled substitution of \ce{Fe^{3+} + Al} for \ce{(Mg, Fe^{2+}) + Si} in bridgmanite \citep[e.g., ][]{Frost2004}. The metallic iron produced in this reaction is higher density and hence migrates to the core, leaving behind oxidized \ce{Fe^{3+}} in the mantle. This reaction thus serves to oxidize the mantle, making $N_b$ less negative. As with reaction (\ref{eq:core_form2}), reaction (\ref{eq:redox_disprop}) will be more efficient in larger planets, due to the greater depth range over which bridgmanite and post-perovskite (\ce{MgSiO3}) \citep[which can similarly incorporate \ce{Fe^{3+}} into its structure; ][]{Catalli2010} are stable. 

It is interesting to note that in a very general sense, Fe disproportionation can be viewed as simply another example of the process of redox gradient formation via gravitational differentiation. The equilibrium constant of (\ref{eq:redox_disprop}) depends on pressure (and hence gravity) via the net volume change of the reaction \citep{ONeill2006}. In situations where disproportionation is favored, the atoms rearrange themselves to minimize Gibbs energy, causing denser metallic iron to sink to the core and the less dense \ce{Fe^{3+}} compounds to remain in the mantle.

Our understanding of the importance of reaction (\ref{eq:redox_disprop}) is still limited by the availability of experimental data. As we show in Section~\ref{sec:crusty}, the upper mantle \ce{Fe^{3+}/Fe^{2+}} ratio needs to reach around 0.3 or more before volcanic outgassing [the $k_2N_b$ term in (\ref{eq:basic1})] becomes a source of net oxidizing power. If in situations where Al is abundant or $p$ and $T$ are high, reaction \eqref{eq:redox_disprop} becomes extremely effective (e.g. on high mass planets), Fe redox disproportionation could contribute significantly to eventual buildup of abiotic \ce{O2} in the atmosphere. {Further experimental study to constrain this issue better in future will be useful}.

\subsection{Initial abundances of \ce{H2} and \ce{H2O} }\label{sec:ICs}

Besides the stellar properties, planet mass and radius, and the mantle FeO abundance, the other key initial conditions required for redox evolution modeling are the volatile layer abundance of \ce{H2} and \ce{H2O}. Hydrogen may be delivered to low mass planets by direct nebular capture \citep{Rafikov2006} or possibly oxidation of metallic iron by \ce{H2O} \citep{Kuramoto1996}. The presence of an initial \ce{H2} envelope is equivalent to starting with an extremely negative value of $N_a$ in equations (\ref{eq:basic1}-\ref{eq:basic2}): it inhibits atmospheric oxidation until all the \ce{H2} is lost to space. Our main aim here is to obtain upper limits on atmospheric \ce{O2} buildup, so in our models we make the assumption that the starting \ce{H2} inventory is negligible.  For \ce{H2O}, we treat the initial abundance as a free parameter varying between 0 and 1~wt\%\footnote{For reference, on Earth the total mass of the surface ocean (1~TO) is 230~ppmw or 0.023~wt\%, while the total mantle \ce{H2O} abundance is somewhere between 0.2 and 13~TO \citep{Hirschmann2009,Marty2012}.}. The densities of planets with \ce{H2O} abundances above a few percent are likely to be sufficiently elevated to allow them to be distinguished from less volatile rich cases \citep{Zeng2013}. 

\section{The upper boundary condition: Atmospheric escape of H}\label{sec:escape}

The final boundary condition we need to incorporate to solve (\ref{eq:basic1}-\ref{eq:basic3}) is the escape term $E$. Atmospheric escape is a complex process that is still incompletely understood. However, of the diverse range of possible atmospheric escape processes, Jeans escape is  {almost} always negligible, while {for the escape of heavy elements such as C and O}, ion-driven processes and sputtering are most important  \citep[e.g., ][]{Lammer2008}. In general, processes driven by the stellar wind appear capable of removing up to tens of bars of gas from planetary atmospheres around G- and M-class stars  \citep{Airapetian2017,Zahnle2017,Dong2017}. These quantities are potentially significant for heavy gases (particularly \ce{N2}; Section~\ref{sec:trap}), but not for \ce{H2} or \ce{H2O}: the equivalent partial pressure of one terrestrial ocean (TO) on Earth is 263~bar. 
Impact-driven escape can be significant \citep{Ahrens1993,Zahnle2017} but does not fractionate gas species. In contrast, extreme ultraviolet (XUV)-driven hydrodynamic escape is capable of removing large quantities of volatiles and always preferentially removes hydrogen as long as it is abundant in the planet's upper atmosphere. For these reasons, it is probably the key process driving redox evolution via escape for planetary atmospheres early in their evolution, and it is what we focus on here.
 
In the absence of any other limits, the ultimate constraint on the rate of XUV-driven escape is the total supply of XUV energy. This leads to the well-known escape rate formula \citep[e.g., ][]{Watson1981,Zahnle1986}
\begin{equation}
\phi_E  = \frac{\epsilon F_{{XUV}}}{4 V_{pot}}\label{eq:Elim}
\end{equation}
where $\phi_E$ is a mass flux [kg/m$^2$/s], $F_{{XUV}}$ is the stellar flux in the XUV wavelength range suitable for ionizing hydrogen ($\sim$10-91~nm) and $V_{pot}=GM_p/ r_p$ is the gravitational potential at the base of the escaping region, with $G$ the gravitational constant and $M_p$ and $r_p$ the planetary radius and mass, respectively. $\epsilon$ is an efficiency factor, which we discuss further in Section~\ref{sec:eff_factor}.  

\begin{figure}[h]
	\begin{center}
\ifarxiv
		{\includegraphics[width=2.65in]{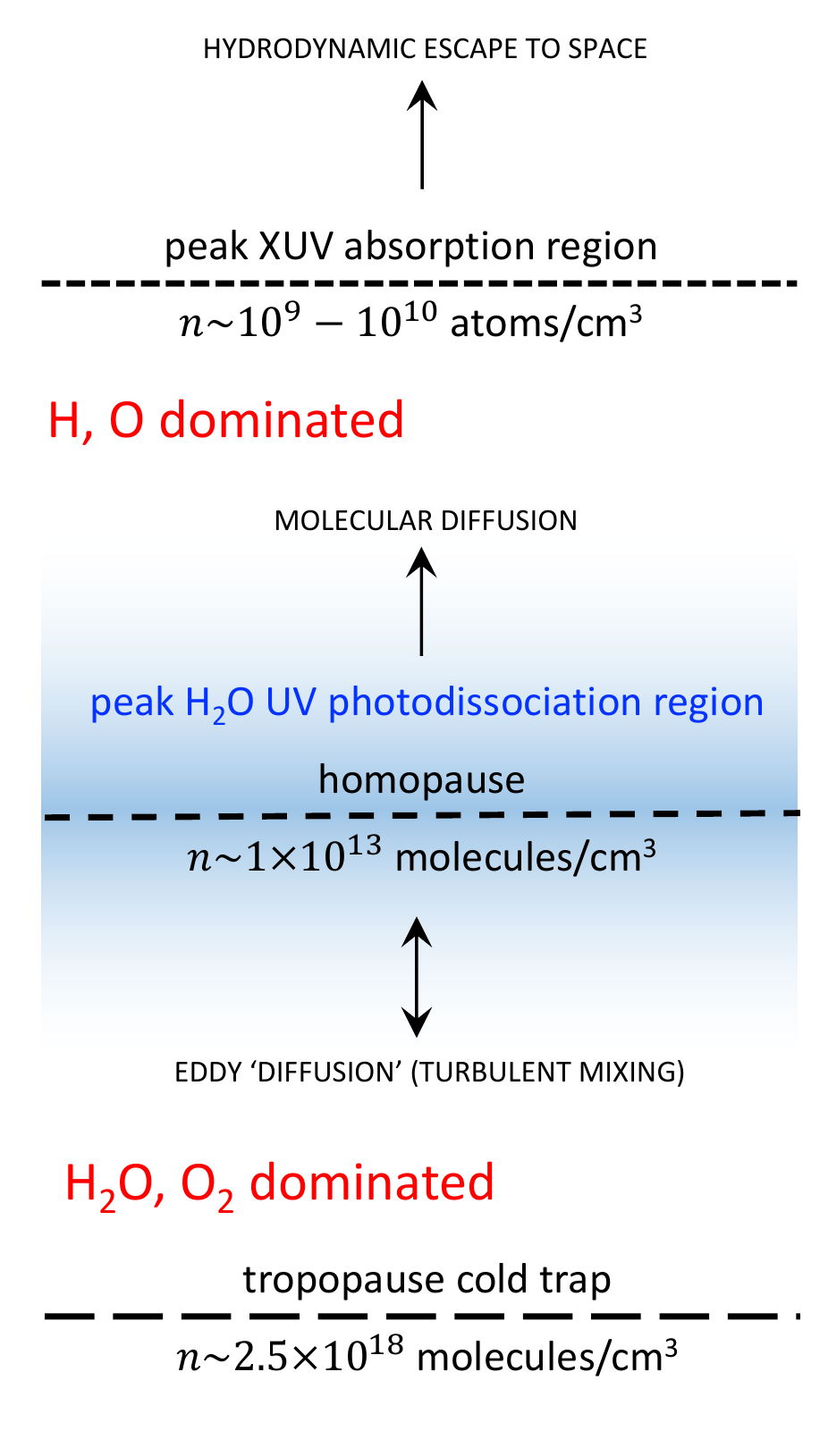}}
\else
		{\includegraphics[width=2.65in]{figures/atm_schematic_v5.pdf}}
\fi
	\end{center}
	\caption{Schematic showing the key regions in a planetary atmosphere undergoing oxidation via \ce{H2O} photolysis and hydrogen loss to space. The approximate number density of each region is also indicated.}
\label{fig:atm_schematic}
\end{figure}

The upper portions of planetary atmospheres may be hydrogen-rich due to the presence of \ce{H2} from volcanic outgassing or a primordial envelope, in which case the oxidation rate is simply $E=4\pi r_p \phi_E / m_p$, where $m_p$ is the proton mass. However, if \ce{H2} is not present, further oxidation can only occur via the photolysis of hydrogen-bearing molecules, of which \ce{H2O} is the most important. Then, the extent to which \ce{H2O} is cold-trapped in the deep atmosphere, the rate at which it is photolyzed in the upper atmosphere, and the rate at which hydrogen diffuses through the homopause all become important (Figure~\ref{fig:atm_schematic}). Cold-trapping is important in a planet's later stages of evolution, once the surface has solidified, and we reserve discussion of it until Section~\ref{sec:crusty}. Diffusion and photochemistry are modeled in the next section, while the escape efficiency is constrained in Section~\ref{sec:eff_factor}.  

\subsection{Diffusion and atmospheric photochemistry}\label{sec:photolim}

During XUV-driven hydrodynamic escape, the composition of the escaping gas, and hence the rate of oxidation of the planet, depends critically on the rate at which products from UV photolysis occurring deeper in the atmosphere diffuse upwards. When diffusion is efficient, the dominant escaping species will be H, with heavier atoms dragged along to an extent that depends on the total flux. In the limit when diffusion is extremely slow or when photolysis products are efficiently recycled, preferential escape of H could be choked off, and net oxidation of the planet would not occur. In this section we model upper atmosphere diffusion and photochemistry to elucidate this critical part of the planetary oxidation problem. Readers not interested in the details should skip to Section~\ref{sec:TPO}, where we summarize the main results.

Water is photolyzed by UV radiation of wavelength $<195$~nm via a number of reactions, the most important of which is 
\begin{equation}
\ce{H2O + h\nu\to OH + H}. \label{eq:photochem1}
\end{equation}
Once atomic H is liberated, it may either react with other atmospheric species or escape to space. When the escape flux is low, H will escape alone, but once it exceeds a critical value, heavier species will also be dragged along. Given a total mass flux $\phi$, the number flux [atoms/m$^2$/s] of a light species $\Phi_1$ and a heavy species $\Phi_2$ per unit surface area can be calculated as a function of their molar concentrations $x_i$ and atomic/molecular masses $m_i$ as 
\begin{eqnarray}
\Phi_1 \approx \left\{
  \begin{array}{lr}
    \phi/m_1 & :  \phi  < \phi_c  \\
    \left[x_1\phi  +  x_1 x_2  (m_2 - m_1) \Phi_{d,2}\right]/ {\overline m}   & :   \phi  \ge \phi_c
  \end{array} \right.
      \label{eq:num_flux_1}
\end{eqnarray}
and
\begin{eqnarray}
\Phi_2 \approx \left\{
  \begin{array}{lr}
   0 & :  \phi  < \phi_c  \\
    \left[x_2\phi +  x_1 x_2(m_1-m_2)  \Phi_{d,1}\right]/ {\overline m}   & :  \phi  \ge \phi_c
  \end{array} \right.
      \label{eq:num_flux_2}
\end{eqnarray}
Here  $\overline m = m_1 x_1 + m_2 x_2$ is the mean molecular/atomic mass of the flow. The $\Phi_{d,i}$ are \emph{equivalent diffusion fluxes}, defined as 
\begin{equation}
\Phi_{d,i} \equiv \frac{b}{H_i}  
\label{eq:diff_flux}
\end{equation}
where $b$ is the binary diffusion coefficient for the two species and $H_i$ is the effective scale height of species $i$ at the base of the escaping region. The quantity $\phi_{c}$ is the critical mass flux required to initiate drag of the heavy species 2 along with the light species 1. It is defined as 
\begin{equation}
\phi_c = \Phi_{d,1}  x_1 \left(m_2 - m_1\right)= \frac{b x_1}{H_1}\left(m_2 - m_1\right).\label{eq:cross_flux}
\end{equation}
This result is easily derived from \eqref{eq:num_flux_2} by noting that the two definitions of  $\Phi_2$ must equal each other when $\phi=\phi_c$, and using the scale height definition \mbox{$H_i = k_BT/m_i g$}, where $g$ is gravity, $k_B$ is Boltzmann's constant and $T$ is temperature.  The familiar expression for diffusion-limited escape of a light minor species through a heavier, non-escaping species simply corresponds to $\Phi_1 = \phi_c / m_1$, or
\begin{equation}
\Phi_1 =   {b x_1}\left(H_2^{-1} - H_1^{-1}\right)
\label{eq:diff_lim}
\end{equation}
If species 1 is H and species 2 is \ce{H2O} or \ce{O2} we can write
\begin{equation}
\Phi_{\ce{H},diff} \approx \frac{b x_{\ce{H}}} {H_s}.
\label{eq:diff_lim_H}
\end{equation}
where $H_s$ is the scale height of the background gas. 
Equations \eqref{eq:num_flux_1}-\eqref{eq:cross_flux} are completely equivalent to the `crossover mass' formalism of \cite{Hunten1987} but are considerably more straightforward to work with. Their derivation from first principles is given in Appendix~\ref{apx:diff}. 

The extreme upper limit on the rate of H liberation during photolysis comes from the supply of UV photons to \ce{H2O}.  However, depending on the atmospheric composition, other chemical pathways may remove H rapidly once it is created.  This may be particularly important in an atmosphere that has already begun to build up some \ce{O2}. Classic studies of martian photochemistry  \citep{McElroy1972b,Yung1999} have shown that the three-body reaction 
\begin{equation}
\ce{H + O2 + M \to HO2 + M},\label{eq:photochem2}
\end{equation}
where M is a background gas molecule, is a key step in the recycling of H when \ce{O2} is present. Once \ce{HO2} has formed, combination with the OH radical
\begin{equation}
\ce{OH + HO2  \to H2O + O2}.\label{eq:photochem3}
\end{equation}
closes the cycle, leading to stabilization of \ce{H2O} against photolysis. On present-day Mars, which has an \ce{H2O}-poor upper atmosphere, this means that hydrogen escape depends on a minor pathway to form \ce{H2}, and H escape is regulated until O and H escape in a 1:2 ratio. However, Mars' crust appears to have oxidized extensively relative to its mantle \citep{Wadhwa2001}, which provides a hint that reaction \eqref{eq:photochem2} may not effectively limit H escape under all circumstances.

To understand the relative importance of chemical and diffusive effects, we have performed simulations using a one-dimensional photochemical model \citep{Wordsworth2016}. Our model calculates the number density time evolution for a given species $n_i$ via the equation
\begin{equation}
\frac{\partial n_i}{\partial t} +  \frac{\partial \Phi_i}{\partial z} =  P_i - L_i 
\label{eq:photochem_master}
\end{equation}
Here $n_i$ is the number density of species $i$ and $P_i$ and $L_i$ are the rates of chemical production and loss. $\Phi_i(z)$ is the number flux due to transport processes. It is defined here as 
\begin{equation}
\Phi_i = -Kn\frac{\partial}{\partial z}\left(\frac{n_i}{n} \right) - D  n_{i,e} \frac{\partial}{\partial z}\left( \frac{n_i}{n_{i,e}}  \right)
\label{eq:photochem_master2}
\end{equation}
where $n_{i,e}\propto \E^{-z/H_i}$, $n\propto \E^{-z/H}$, $H$ is the mean scale height of the atmosphere, $K$ is the eddy diffusion coefficient and \mbox{$D=b/n$}, with $b$ is the binary molecular diffusion coefficient and $n$ the total number density \citep{Yung1999}. We treat $K$ as constant with height, with a nominal value of $10^5$~cm$^2$/s. Our representation of $b$ is summarized\footnote{The ``all others" category in Table~\ref{tab:binary_diff} uses data for \ce{O2} in \ce{H2O}. Although data is not available for every possible interacting pair, differences between $b$ values among species are small in general. Because we are most interested in order of magnitude changes to escape rates,  our use of a reduced set of $b$ values here is unlikely to have a significant impact on our results. }  in Table~\ref{tab:binary_diff}. 

\begin{table}[h]
\centering
\begin{tabular}{cc}
\hline
\hline
Species & $b$ [molecules/cm/s]   \\
\hline
\ce{O}, \ce{O(^1D)} in \ce{H2O} & $1.06\times10^{17} T^{0.774}$      \\
\ce{H} in \ce{H2O} & $6.6\times10^{17} T^{0.7}$    \\
\ce{H} in \ce{O}      & $4.8\times10^{17} T^{0.75}$     \\
\ce{H2} in \ce{H2O}& $2.7\times10^{17} T^{0.75}$    \\
all others & $1.37\times10^{16} T^{1.072}$      \\
\hline
\hline
\end{tabular}
\caption{Values of the binary diffusion coefficient $b$ used in the photochemical and escape calculations for various interacting species. Data for all species were taken from \cite{Zahnle1986} and cross-checked vs. data in \cite{Marrero1972}. 
}
\label{tab:binary_diff}
\end{table}

\begin{figure}[h]
	\begin{center}
\ifarxiv
		{\includegraphics[width=3.25in]{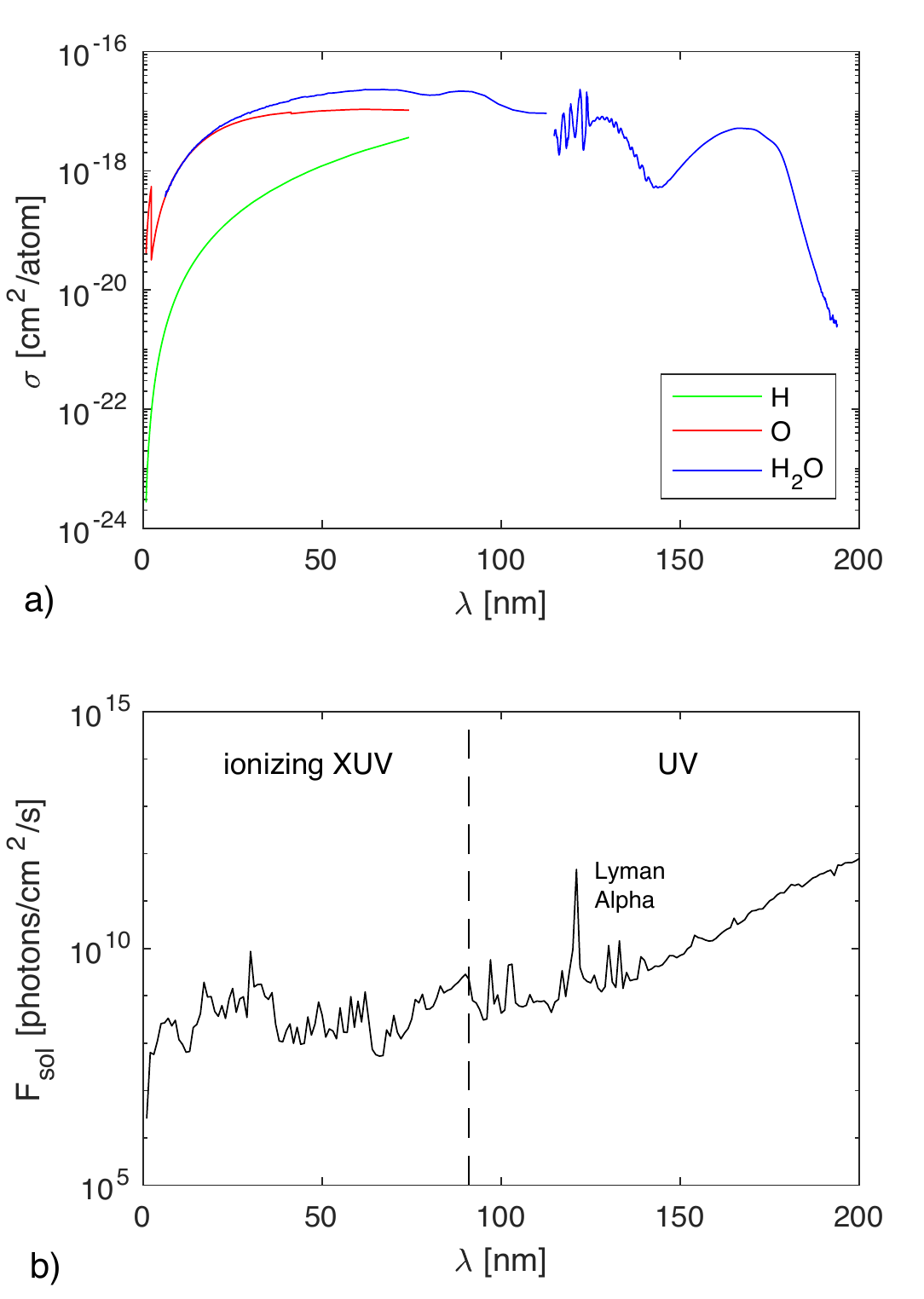}}
\else
		{\includegraphics[width=3.25in]{figures/cross_secs_stel_specs_v2.pdf}}
\fi
	\end{center}
	\caption{a) Photoionization cross-sections for H and O and photodissociation cross-section for \ce{H2O} as a function of wavelength, based on data from \cite{Yeh1985}, \cite{Yeh1993}, \cite{Chan1993} and \cite{Mota2005}. b) Present-day solar flux at Earth orbit across the same wavelength range, based on data from \cite{Thuillier2004}. }
\label{fig:cross_secs_stel_specs}
\end{figure}

Photodissociation reaction rates are calculated as 
\begin{equation}
J_k(z) = \frac 14 \int_{\lambda_1}^{\lambda_2} Q_k(\lambda)\sigma_k(\lambda)F_{{UV}}(z,\lambda)d\lambda
\end{equation}
where $\lambda$ is wavelength, $\sigma_k$ and  $Q_k$ are the absorption cross-section and quantum yield of photoreaction $k$, respectively, and $F_{UV}$ is the incoming stellar UV flux at wavelengths below $\lambda_2 = 195$~nm. The nominal spectrum for $F_{UV}$ is shown in Figure~\ref{fig:cross_secs_stel_specs}; see Section~\ref{sec:TPO} for a discussion of our treatment of M-star UV spectra and temporal evolution. The factor of $1/4$ accounts for day-night averaging and the mean angle of propagation (assumed to be $60^\circ$ here). In the nominal simulations, we allow both UV and XUV radiation to contribute to photolysis\footnote{We tested the effects of removing all the XUV radiation used to power escape first, and found that the influence on our results was insignificant.}, setting $\lambda_1 = 1$~nm. $F_{UV}(z,\lambda)$ is calculated in each layer from the number density and total absorption cross-section assuming a mean propagation angle of 60$^\circ$. The average value of $J_k(z)$ is then used when solving (\ref{eq:photochem_master}).  

We solve this coupled system of equations for 10 chemical species and 50 vertical layers using an adaptive timestep semi-implicit Euler method, with the reaction rate coefficients given in Table~4. 
The calculation is continued until a steady state is reached, which we check by observing the time evolution of all species at the top and bottom boundaries of the model. Our diffusion scheme, which is based on a weighted centered finite difference, has been tested vs. analytic results and verified to conserve molecule number to high precision.

{In previous studies focused on abiotic oxygen production in Earth-like atmospheres, strong emphasis has been placed on the ability of photochemical models to satisfy \emph{redox balance}  \citep[e.g., ][]{Domagal2014,Harman2015a}, usually defined simply as ``conservation of free electrons'' \citep{Harman2015a}. This emphasis is important for problems where the total oxidation state of the atmosphere + oceans $N_a$ is assumed to remain constant with time. Because our photochemical model conserves atom number to high precision, it also conserves free electrons. 
This \emph{does not} mean that the number of free electrons in the atmosphere necessarily remains constant before equilibrium is reached in our simulations, as we allow for the possibility of O and H fluxes through the top and bottom model boundaries, which evolves the atmospheric composition. However, our coupled approach to redox flow (Fig.~\ref{fig:box_model_schematic}) means that in later sections when we link the atmosphere with the planetary interior, the global number of accessible electrons $N_e = -(N_a + N_b + N_c)$ is only altered by the escape to space term $E(t)$. 
Hence our model satisfies redox balance according to the standard definition. Importantly, it has the additional advantage of requiring no ad-hoc assumptions about the redox state of the mantle after the initial conditions have been set. }

The {photochemical} model domain is defined from $10^4$~Pa at the base to $10^{-5}$~Pa at the top, to encompass the entire range over which \ce{H2O} photolysis is important. We have confirmed that our results are insensitive to increases in this pressure range. We initialize the atmosphere with constant molar concentrations of \ce{H2O} and \ce{O2}, and the abundances of all other species set to zero. The default boundary condition is zero flux (Neumann) at the top and bottom of the model. For \ce{H2O} and \ce{O2}, we use Dirichlet boundary conditions at the bottom of the model to keep their molar concentrations fixed. For \ce{H}, we force the molar concentration gradient to be zero at the top boundary, corresponding to diffusion-limited escape. 

\begin{figure}[h]
	\begin{center}
\ifarxiv
		{\includegraphics[width=3.5in]{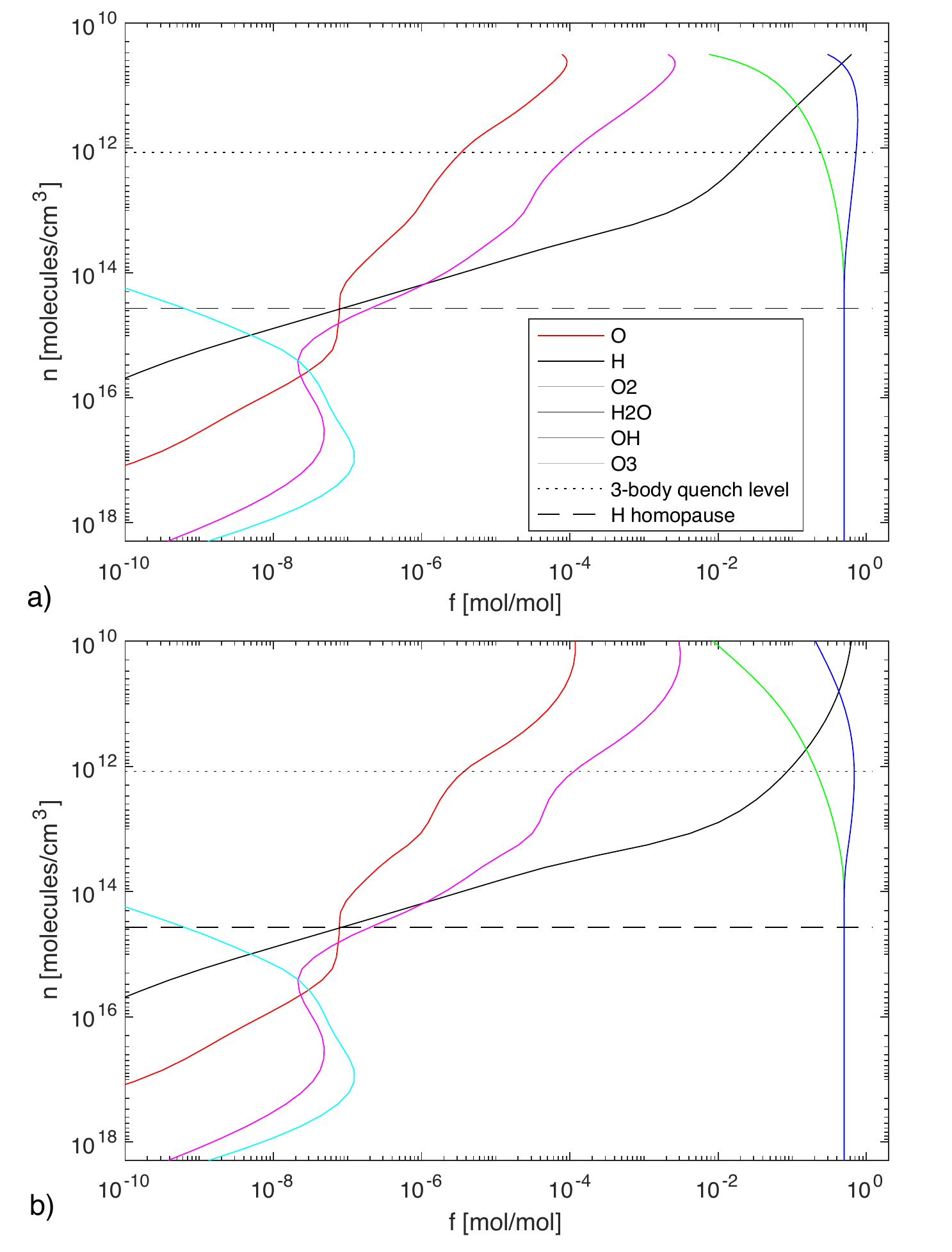}}
\else
		{\includegraphics[width=3.5in]{figures/photochem_example_1D_v8.pdf}}
\fi
	\end{center}
	\caption{Example output from the one-dimensional photochemical model with a) fixed and b) varying total number density and mean scale height. The dashed and dotted lines show the locations of the homopause (for H) and the three-body quench level defined by (\ref{eq:quench}), respectively. In both cases, the H escape rate is $1\times10^{12}$~atoms/cm$^2$/s. }
\label{fig:photochem_example}
\end{figure}

Figure~\ref{fig:photochem_example} shows the results of an example calculation with Earth-like UV and XUV insolation, $K=10^5$~cm$^2$/s, and \ce{O2} molar concentration of $0.5$~mol/mol at the base of the domain.  The atmospheric composition is dominated by \ce{H2O}, \ce{O2}, \ce{OH} and \ce{H}, with \ce{O} and \ce{O3} existing as minor constituents near the top and base of the domain, respectively. 
Here, the H escape rate is $\Phi_{\ce{H},diff} = 8.3 \times 10^{11}$~atoms/cm$^2$/s. For comparison, the extreme upper limit on the \ce{H2O} photolysis rate is the total accessible UV photon flux 
\begin{equation}
\Phi_{\ce H,{UV}} = \frac 14 \int_{\lambda_1}^{\lambda_2} F_{UV}(z,\lambda)d\lambda.\label{eq:UVphotolimit}
\end{equation}
For present-day Earth, \mbox{$\Phi_{\ce H,{UV}} \approx 1.9\times10^{12}~\mbox{atoms/cm$^2$/s}$}, or about 15 times larger than the energy-limited escape rate for hydrogen atoms, $\Phi_{\ce H,{E}} = m_p\phi_E$. Obviously, both $\Phi_{\ce H,{UV}}$ and $\Phi_{\ce H,{E}}$ vary with the incident stellar flux.

As can be seen, there is a sharp decline in the concentration of H below a given depth, due to the rapid increase in the rate of reaction (\ref{eq:photochem2}) with depth. Because (\ref{eq:photochem2}) occurs at a rate \mbox{$\partial n_{\ce{H}}/\partial t = -k_3 n n_{\ce{O2}} n_{\ce{H}}$}, with $k_3$ defined by C3 in Table~4, the total number density $n_b$ at which this transition occurs in equilibrium can be estimated as
\begin{equation}
n_b^2 \sim \frac{1}{k_3x_{\ce{O2}}\tau_{diff}} \label{eq:quench}
\end{equation}
with $x_{\ce{O2}}$ defined at $n_b$ and $\tau_{diff} = H_s^2 /(K+D)$ a characteristic timescale for  diffusion of H. Equation~(\ref{eq:quench}) is easily solved in general, but for situations where the transition occurs above the homopause, as in Fig.~\ref{fig:photochem_example}, it can be simplified further to
\begin{equation}
n_b \sim \sqrt[3]{\frac b {k_3x_{\ce{O2}}H_s^2}}.
\label{eq:quench2}
\end{equation}
If we treat the atmosphere above $n_{b}$ as a single region, the overall H budget can be approximated as a balance between H production and loss. {Production comes from \ce{H2O} photolysis, while loss must occur through downwards eddy diffusion, because molecular diffusion preferentially transports hydrogen upwards. Hence}
\begin{equation}
\int_{z_b}^\infty P_{\ce{H}} dz \approx \Phi_{\ce{H},diff} + \frac{n_{\ce{H}} H_s}{\tau_{diff,K}}.
\end{equation}
Here $z_b$ is the altitude at which $n=n_b$,  \mbox{$P_{\ce{H}}\approx J_{\ce{H2O}}n_{\ce{H2O}}$}, and $\tau_{diff,K} = H_s^2/K$ is the timescale for eddy diffusion of H downwards into the lower atmosphere. Assuming $x_{\ce{O2}}\approx1-x_{\ce{H2O}}$ and making use of the diffusion-limited escape equation (\ref{eq:diff_lim_H}), we can rearrange to get 
\begin{equation}
\Phi_{\ce{H},diff} \approx  \frac{\int_{z_b}^\infty J_{\ce{H2O}}n_{\ce{H2O}} dz}{1 +  K n_b / b }.\label{eq:quench_analytic}
\end{equation}

\begin{figure}[h]
	\begin{center}
\ifarxiv
		{\includegraphics[width=3.5in]{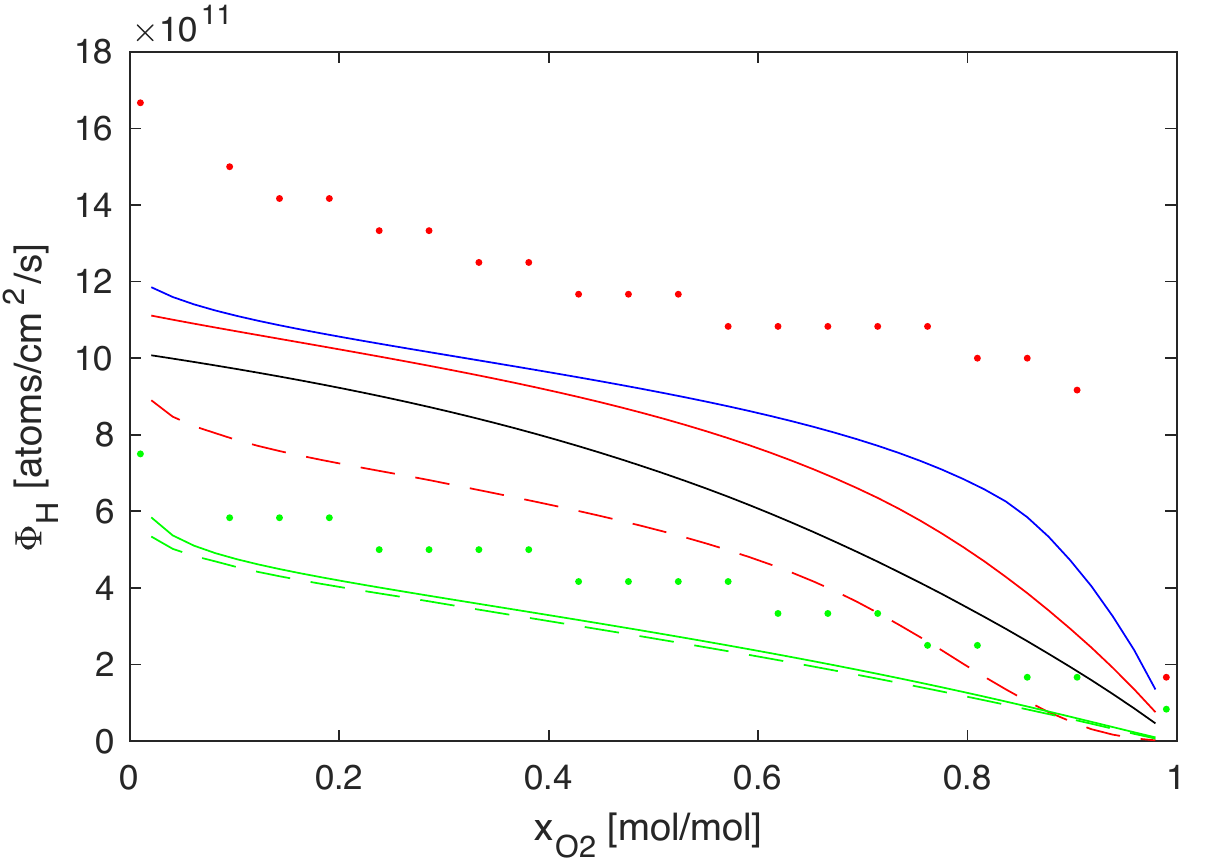}}
\else
		{\includegraphics[width=3.5in]{figures/photochem_flux_1D_v4.pdf}}
\fi
	\end{center}
	\caption{Diffusion-limited H escape rate $\Phi_{\ce{H}}$ as a function of the \ce{O2} molar concentration of the bulk atmosphere $x_{\ce{O2}}$. Black, red and blue solid lines show results for $K=10^4,10^5$ and $10^6$~cm$^2$/s, respectively, while the green solid line shows the results for $K=10^5$ with all reactions removed expect A4 and C3 (see Table~4). The dashed lines show the semi-analytic result (\ref{eq:quench_analytic}) for the two \mbox{$K=10^5$~cm$^2$/s} cases. In both cases, the production term $\int_{z_b}^\infty J_{\ce{H2O}}n_{\ce{H2O}} dz$ is derived from the model results. {Finally, the dots show H escape upper limits estimated from a code version with varying $n$ and $H_s$ (see main text).}}
\label{fig:photochem_flux_1D}
\end{figure}

Figure~\ref{fig:photochem_flux_1D} compares this result with the H loss rate calculated by the model as a function of the base \ce{O2} molar concentration. As can be seen, in both cases the escape rate of hydrogen steadily decreases as \ce{O2} builds up in the atmosphere. The analytic prediction does a reasonable job despite its simplicity, indicating that we have captured the key features of the one-dimensional model. The small systematic underprediction of the model results is due to the neglect of additional O-H reactions that recycle H after interaction with \ce{O2}, as is clear from the intercomparison with all additional reactions removed (green lines). While there are variations, to a first approximation the decrease in $\Phi_{\ce{H}}$ is  linear with $x_{\ce{O2}}$. 

Importantly, \ce{O2} molar concentrations below a few percent do not significantly decrease H escape below the \ce{O2}-free value. The key reason for this is that the {3-body reaction} (\ref{eq:photochem2}) only dominates H removal relatively deep in the atmosphere. Like hydrogen balloons released from an aeroplane, hydrogen atoms liberated from \ce{H2O} above both the homopause and the $n_b$ level mainly escape upwards to space, rather than mixing downwards. On present-day Mars, the upper atmosphere is extremely poor in \ce{H2O}, and most photolysis occurs deeper in the atmosphere. Recent analyses of the martian atmosphere suggest that loss of atomic H is enhanced when water is able to propagate to the high atmosphere \citep{Chaffin2017}. Our results are consistent with this prediction.

\begin{figure}[h]
	\begin{center}
\ifarxiv
		{\includegraphics[width=3.5in]{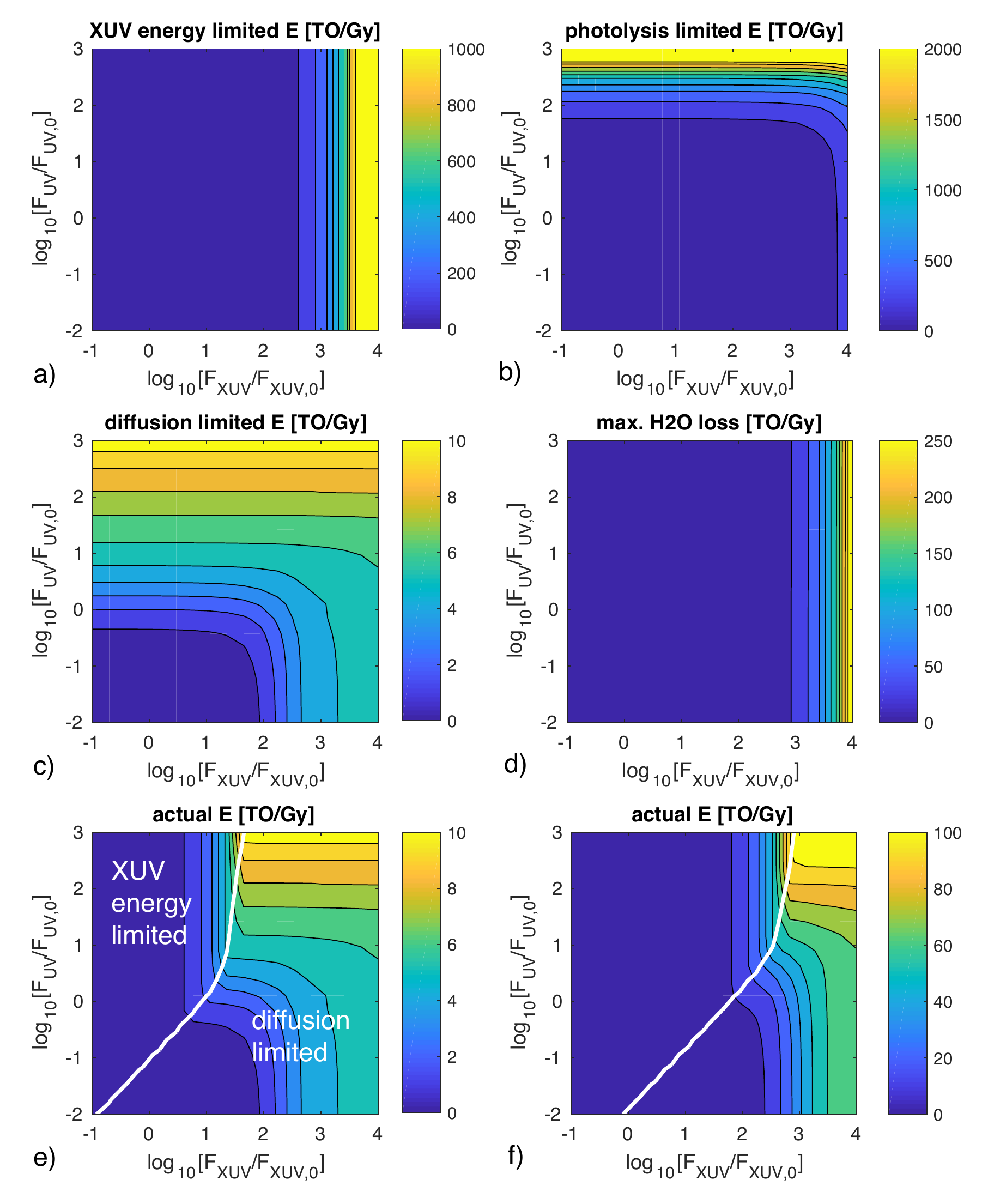}}
\else
		{\includegraphics[width=3.5in]{figures/photo_eff_v5.pdf}}
\fi
	\end{center}
	\caption{Escape rates as a function of XUV 
	and UV 
	stellar fluxes incident at the top of the atmosphere. Plots a-c) show atomic H escape rate ($E$) from a 100\% \ce{H2O} atmosphere assuming that the limiting factor on escape is a) the supply of XUV energy to power escape  (equation~\ref{eq:Elim}), b) the rate of supply of UV and XUV photons (equation~\ref{eq:UVphotolimit}), or c) the upwards diffusion of H through the homopause (equation~\ref{eq:diff_lim}), respectively. Panel d) shows the \ce{H2O} loss rate assuming that excess XUV energy is used to power O escape. Panel e) shows the actual escape rate of H obtained by combining the limits in b) and c), with divisions between the regimes indicated by the white lines. Panel f) shows the same thing for a 10~$M_E$ super-Earth.}
\label{fig:phot_eff}
\end{figure}

{In most simulations, we did not force the total number density $n$ to evolve with time. We regarded this as an acceptable approximation because the escape rate of H is set at either the homopause or at $n_b$, where it is a minor constituent that does not significantly modify $n$. Nonetheless, as a check on our results, we performed some simulations where $n$ and the mean scale height $H_s$ were allowed to evolve with time in the governing equations. To maintain model stability in these simulations, it was necessary to use a separate Crank-Nicholson scheme for the diffusion solver and a small timestep, resulting in longer simulation times. Fig.~\ref{fig:photochem_example}  compares simulations with fixed and varying background profiles for the same boundary conditions. As can be seen, the only significant variation to molar concentrations occurs at the very lowest number densities, well above the $n_b$ level. We also performed $n$ and $H_s$-varying simulations where we increased $\Phi_{\ce{H}}$ until $f_{\ce{H}}$ became negative at the top of the atmosphere, and recorded the last stable value for $\Phi_{\ce{H}}$. The resulting upper limits on $\Phi_{\ce{H}}$ are displayed as dots in Fig.~\ref{fig:photochem_flux_1D}. As can be seen, they are within a factor of 1.5 or less of the standard diffusion limits for $\Phi_{\ce H}$ for most values of $x_{\ce {O2}}$. Selected tests at high UV fluxes (not shown) showed similar behaviour. Based on this, we decided to keep the standard model setup for our main calculations. The sensitivity of the model results to a twofold increase in H escape rates is discussed in Section~\ref{sec:melty}.}

Having established the effect of \ce{O2} on H escape under Earth-like XUV and UV conditions, we now explore a wider range of stellar fluxes. Figure~\ref{fig:phot_eff} shows the results of simulations where we varied stellar XUV and UV separately over four orders of magnitude. For both XUV and UV, we use axes scaled to Earth's present-day averaged fluxes ($F_{{XUV},0}=4\times10^{-3}$~W/m$^2$) and  ($F_{{UV},0}=1.9\times10^{12}$~photons/cm$^2$/s), 
as calculated by integrating the solar spectrum data of \cite{Thuillier2004}. The quantity plotted is the oxidation rate $E$ (see~Figure~\ref{fig:box_model_schematic}) in TO/Gy. We calculate this as 
\begin{equation}
E = \frac{4\pi r_P^2  t_{\mbox{Gy}} }{N_{e,\mbox{TO}}}\Phi_{\ce{H}}\label{eq:Edefn}
\end{equation}
where $\Phi_{\ce{H}}$ is the number flux in question ($\Phi_{\ce{H},E}$, $\Phi_{\ce{H},diff}$ or $\Phi_{\ce{H},UV}$) and $t_{Gy}$ is the number of seconds in 1~Gy. Figs.~\ref{fig:phot_eff}~{a) and b)} show the XUV energy and photolysis limits on escape we have already discussed. The XUV energy limit is a linear function of $F_{{XUV}}$ only, while the photolysis limit is a linear function of both $F_{{XUV}}$ and $F_{{UV}}$. Fig.~\ref{fig:phot_eff}~{c)} shows the actual diffusion-limited H escape rate obtained from the one-dimensional photochemical model.

Figure~\ref{fig:phot_eff}e) shows the actual H escape rate, which we obtain by combining all these three limits. As has been suggested in earlier work \citep{Wordsworth2013c,Schaefer2016}, we find that the photolysis limit is never reached in practice. Instead, XUV energy-limited escape transitions to H diffusion limited escape at $F_{{XUV}}$ levels between around 10 and 30 times present-day Earth. The maximum oxidation rate obtained in the 100\% pure \ce{H2O} atmosphere is around 10~TO/Gy. In similar simulations performed with a 10\% \ce{H2O}, 90\% \ce{O2} atmosphere (not shown), the maximum escape rate was approximately 1~TO/Gy, confirming that the quasi-linear dependence of $\Phi_{H,diff}$ on  $x_{\ce{O2}}$ seen in Figure~\ref{fig:photochem_flux_1D} holds across the range of stellar fluxes studied. Figure~\ref{fig:phot_eff}f) is the same as Figure~\ref{fig:phot_eff}e) but for a 10~$M_E$ super-Earth. As can be seen, escape is energy-limited {over a wider range of fluxes} in this case, because the higher super-Earth gravity makes interspecies diffusion much more effective.

We have also performed photochemical simulations with diffusion-limited escape of O, \ce{H2} and \ce{OH} included, and found that this has little effect on the diffusion-limited H escape rate. \ce{H2} and \ce{OH} are not abundant enough at the top of the atmosphere in the simulations to affect redox evolution significantly when they escape. O is a major atmospheric species at high UV and XUV levels, but its diffusion through \ce{O2} and \ce{H2O} does not appear to strongly affect the H diffusion rate.

The peak values of $E$ in Fig.~\ref{fig:phot_eff} are close to the maximum rate at which the planet can oxidize via H loss, because if O is effectively dragged along with the H atoms this will make the escaping gas more oxidizing and hence counteract increases in $E$. Indeed, for the idealized case of pure escaping H (species 1) and O (species 2) with $x_1=x_{\ce{H}}=2/3$, $x_2=x_{\ce{O}}=1/3$, $m_1 = 1$~amu and $m_2 = 16$~amu, the number flux in (\ref{eq:Edefn}) becomes
\begin{equation}
\Phi_{\downarrow,2} \approx \frac{5b m_p g}{k_B T},\label{eq:phidownarrow}
\end{equation}
where $m_p$ is the proton mass, $g$ is gravity, $k_B$ is Boltzmann's constant and $T$ is temperature (see Appendix~\ref{apx:diff}). This expression yields $E = 21.6$~TO/Gy for Earth with $T = 300$~K, or about twice the maximum value in Fig.~\ref{fig:phot_eff}. The difference is mainly due to the lower values of $x_{\ce{H}}$ that occur when a full photochemical calculation is performed, because some liberated H is always mixed downwards into the lower atmosphere.

While escape of O along with H cannot significantly alter $E$, it will still contribute to the overall rate of water (\ce{H2O}) loss. Water loss is  important both from a habitability perspective, and because it plays a key role in regulating a planet's magma ocean phase (next section). The lower limit on \ce{H2O} loss is simply the value of $E$ in Fig.~\ref{fig:phot_eff}. The upper limit can be estimated by assuming that O also escapes, in the diffusion-limited regime, at a rate determined by the excess XUV energy available to power escape. This loss rate is shown in Fig.~\ref{fig:phot_eff}d. As can be seen, loss rates of 10s of TO per Gy are theoretically possible at the highest XUV fluxes studied.

To summarize, the key conclusions of this subsection are a) the rate of planetary oxidation via \ce{H2O} photolysis and H escape $E$ is either XUV energy limited or diffusion limited, depending on the relative XUV and UV fluxes and b) the decrease in the diffusion-limited H escape rate as the \ce{O2} abundance increases is approximately linear. 

\subsection{Escape efficiency} \label{sec:eff_factor}

Having assessed the role of photolysis and diffusion in the transport of H to the base of the hydrodynamic escape region, we now analyze the efficiency of the escape process itself. The energy-limited hydrodynamic escape equation (\ref{eq:Elim}) is useful because of its extreme simplicity. This simplicity comes at a cost: all information on conduction and radiative transfer is subsumed into the efficiency factor ($\equiv$ fudge factor) $\epsilon$. Because of the range of processes we are already incorporating in this study, we leave development of a rigorous model of multi-species hydrodynamic escape to future work. However, we can still understand the range of possibilities for H escape by studying limiting cases for the behavior of $\epsilon$ as a function of time.

Physically, we should expect that radiative processes will be more important (and hence $\epsilon$ will be lower) in situations where a) more radiating species are present or b) temperatures are high enough to make new types of emission effective. For pure hydrogen, previous work on hot Jupiters has shown that the main sources of radiation are Lyman-$\alpha$ radiative cooling and vibrational transitions of the \ce{H_3^+} molecule \citep[e.g., ][]{Yelle2004,Murray2009}. Lyman-$\alpha$ cooling begins to dominate at escape temperatures around $10^4$~K, which is above the blowoff temperature\footnote{Blowoff occurs when the atmospheric scale height approaches the planet radius, or $ T \approx r_p m_p g/k_B$ for an H-dominated flow.} on terrestrial-mass planets for a pure atomic H flow, but not when both O and H escape. \ce{H_3^+} emission is important only in atmospheres where \ce{H2} is already abundant. The role of heavier ions in the radiative transfer of an escaping flow is still poorly understood. However, both NLTE emission from the vibration-rotation bands of molecules such as \ce{CO2} and electronic transitions associated with N, O and C ions (airglow) are likely to be important.

Here we study three scenarios for $\epsilon$. In the first ($\epsilon_1$) we assume that $\epsilon$ maintains a constant high value of 0.3 at all XUV fluxes. In the second ($\epsilon_2$), we assume that $\epsilon = 0.15$ at low XUV levels\footnote{In \cite{Owen2016}, it is argued that because at low XUV levels the exobase on low mass planets will be below the sonic point, escape will proceed extremely slowly, at the Jeans limit, and little hydrogen will be lost. However, \cite{Johnson2013}, who performed sophisticated Monte Carlo simulations that relaxed the continuum fluid assumption, showed that energy limited escape is not dependent on the flow becoming supersonic below the exobase, so this outcome is unlikely to be valid in reality.}, but when XUV is high, radiative effects act to cool the flow. Specifically, we assume that once O atoms begin to be dragged with the escaping H, they cool the flow so effectively that $\phi$ is never allowed to increase above $\phi_c$ (see equation~\ref{eq:cross_flux}). In essence, this leads to close to the same limit as the diffusion-limited H escape in Fig.~\ref{fig:phot_eff}. $\phi_c$ is calculated using  the binary diffusion coefficient for  O and H in Table~\ref{tab:binary_diff} and a homopause temperature of $300$~K. Because $\Phi_{\downarrow,2}\propto b/T = T^{-0.25}$, the sensitivity of our results to the assumed homopause temperature is very low.

For the third case ($\epsilon_3$),  we allow O escape to occur and assume radiative cooling by O is not effective, but we allow for Lyman-$\alpha$ cooling by H atoms. We account for the fact that O drag strongly decreases the scale height of the escaping flow, which means it must heat much more before effective hydrodynamic escape occurs. At these higher temperatures, Lyman-$\alpha$ cooling of the H could potentially become important. We represent the Lyman-$\alpha$ cooling limit in a simple way by approximating the escaping wind as isothermal and the density structure as hydrostatic, following \cite{Murray2009}. We write the escape flux as $\phi = \rho_t c_t$, where $\rho_t$ and $c_t$ are the density and sound speed at the transonic point, respectively. Assuming $T=10^4$~K, $c_t \approx \sqrt{2\times10^4 k_B/\overline m}$, where $\overline m$ is the mean atomic mass of the neutral flow and the factor of two accounts for ionization of all H to \ce{H^+} and O to \ce{O^+}. We neglect  higher ionization states than \ce{O^+}.

From the transonic rule, the transonic point radius is $r_t = GM_P/2c_t^2$ \citep{Pierrehumbert2011BOOK}. If we assume that ionization occurs rapidly near the base of the flow, hydrostatic balance in spherical coordinates allows us to write 
 \begin{equation}
\rho_t \approx \overline m n_{+,base} \E^{ 2\left(1 - {r_t}/{r_p}\right)}
\end{equation}
where $n_{+,base}$ is the number density of H ions at the base. Finally, assuming that the density at the point of peak XUV absorption is determined by ionization equilibrium, we can balance photoionization and radiative recombination to find 
 \begin{equation}
n_{+,base} \approx \sqrt{\frac{F_{{XUV}} }{h\nu_0 H_s \alpha_R}  }.
\end{equation}
Here $h\nu_0 \approx 20$~eV is the mean energy required for one photoionization and \mbox{$\alpha_R = 2.7\times10^{-13}(T/10^4)^{-0.9}$~cm$^3$/atom/s} is the hydrogen Case B radiative recombination coefficient \citep{Spitzer2008}. 
The radiative escape efficiency limit is then calculated as $\epsilon_{Ly-\alpha} = 4V_{pot}\phi/F_{{XUV}}$. To complete our prescription of $\epsilon_3$, we assume that it is never greater than $0.15$ or less than $\epsilon_2$, such that 
\begin{eqnarray}
\epsilon_3 = \left\{
  \begin{array}{lcc}
   \epsilon_2 		   \quad&:&\quad  \epsilon_{Ly-\alpha}  \le \epsilon_2  \\
   \epsilon_{Ly-\alpha}  \quad&:&\quad \epsilon_2<\epsilon_{Ly-\alpha}  < 0.15   \\
   0.15  			   \quad&:&\quad  \epsilon_{Ly-\alpha}  \ge 0.15 .\\
  \end{array} \right.
      \label{eq:esc_eff_3}
\end{eqnarray}

\begin{figure}[h]
	\begin{center}
\ifarxiv
		{\includegraphics[width=3.5in]{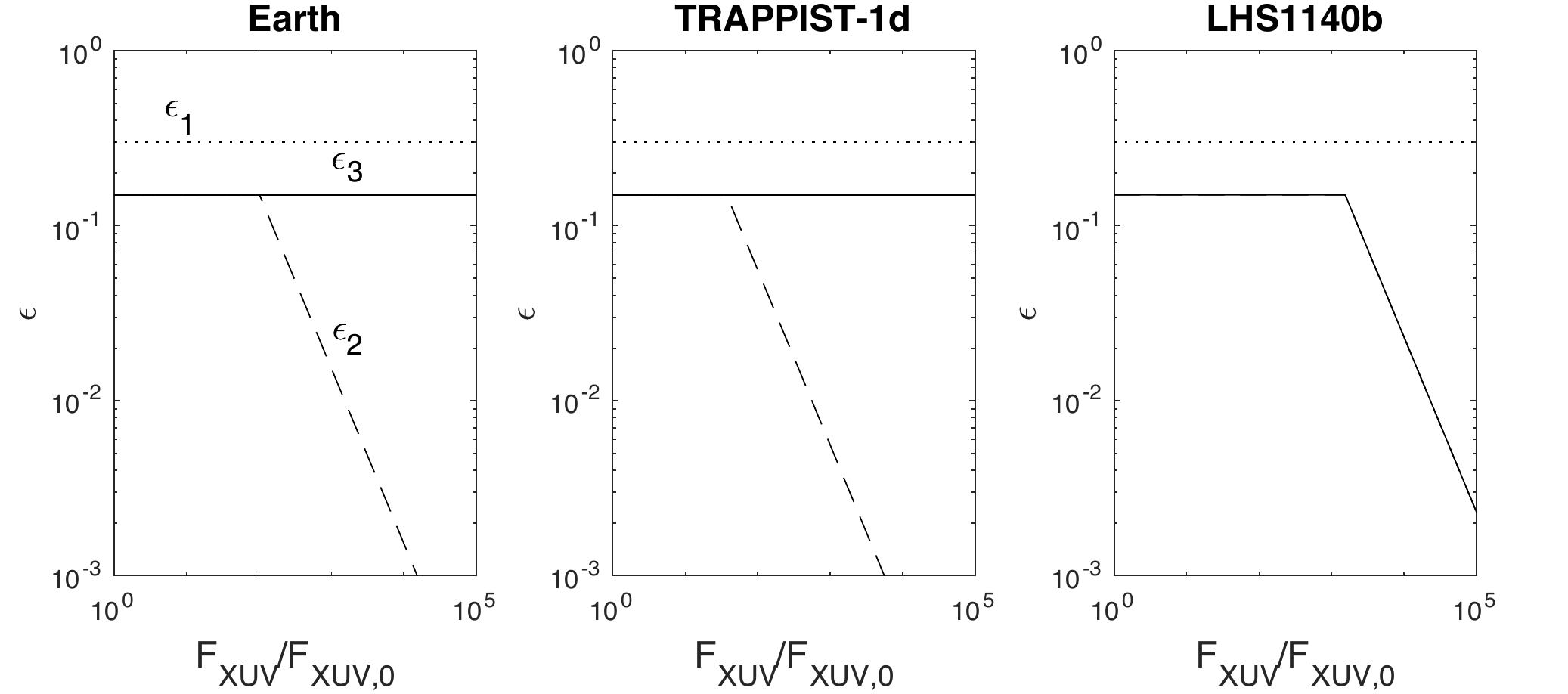}}
\else
		{\includegraphics[width=3.5in]{figures/escape_efficiency_v7.pdf}}
\fi
	\end{center}
	\caption{Plot of the escape efficiency $\epsilon$ as a function of the incoming XUV flux, for Earth ($M_p=M_E$, $r_p=r_E$), {TRAPPIST-1d} (nominal $M_p = 0.41 M_E$, $r_p = 0.772 r_E$) and LHS1140b ($M_p=6.65M_E$, $r_p=1.43r_E$). Here $F_{XUV,0} = 4.0\times10^{-3}$~W/m$^2$ is Earth's present-day received XUV flux. The dotted, dashed and solid lines correspond to $\epsilon_1$, $\epsilon_2$ and $\epsilon_3$ as described in the text. }
\label{fig:esc_eff}
\end{figure}

These three cases for $\epsilon$  are plotted in Figure~\ref{fig:esc_eff} as a function of $F_{{XUV}}$ for three terrestrial-mass planets. For Earth, $\epsilon_2$ decreases rapidly after around 100 times the present-day XUV level, while $\epsilon_3$ does not decline until a flux  of around $10^4 F_{{XUV},0}$ is reached. For the lower density\footnote{Throughout this paper, we use the reported mass and radius values for all exoplanets. 
For the TRAPPIST planets and LHS1140b, in particular, the uncertainties in these values should be borne in mind when interpreting the results.} TRAPPIST-1d, $\epsilon_2$ is lower because O drag commences sooner. Finally, for the super-Earth LHS1140b, the higher gravity enhances diffusive separation of O and H and increases the XUV flux required for O drag to commence to around $2\times10^3  F_{{XUV},0}$. However, the higher gravity also means very high flow temperatures are required for rapid escape once O drag does commence. Hence in our simple model, Ly-$\alpha$ cooling in the mixed H-O flow is predicted to be so efficient that $\epsilon_2=\epsilon_3$ for all $F_{XUV}$. Based on this analysis, we choose to treat $\epsilon_1$ and $\epsilon_2$ as upper and lower limits on escape efficiency, and ignore the complication of implementing Lyman-$\alpha$ cooling directly in our coupled model.

In general, hydrodynamic escape should always become more affected by radiative cooling as the planet mass or the mean molar mass of the flow increases. Effective escape requires the upper atmosphere to be heated until its scale height starts to approach the planetary radius, and this requires much higher temperatures when the planet is massive or the escaping species is heavier than pure H. But higher temperatures frequently lead to more efficient radiative processes, which steal energy that could otherwise be used to power escape.

\subsection{Total potential oxidation rates}\label{sec:TPO}

\begin{figure}[h]
	\begin{center}
\ifarxiv
		{\includegraphics[width=3.0in]{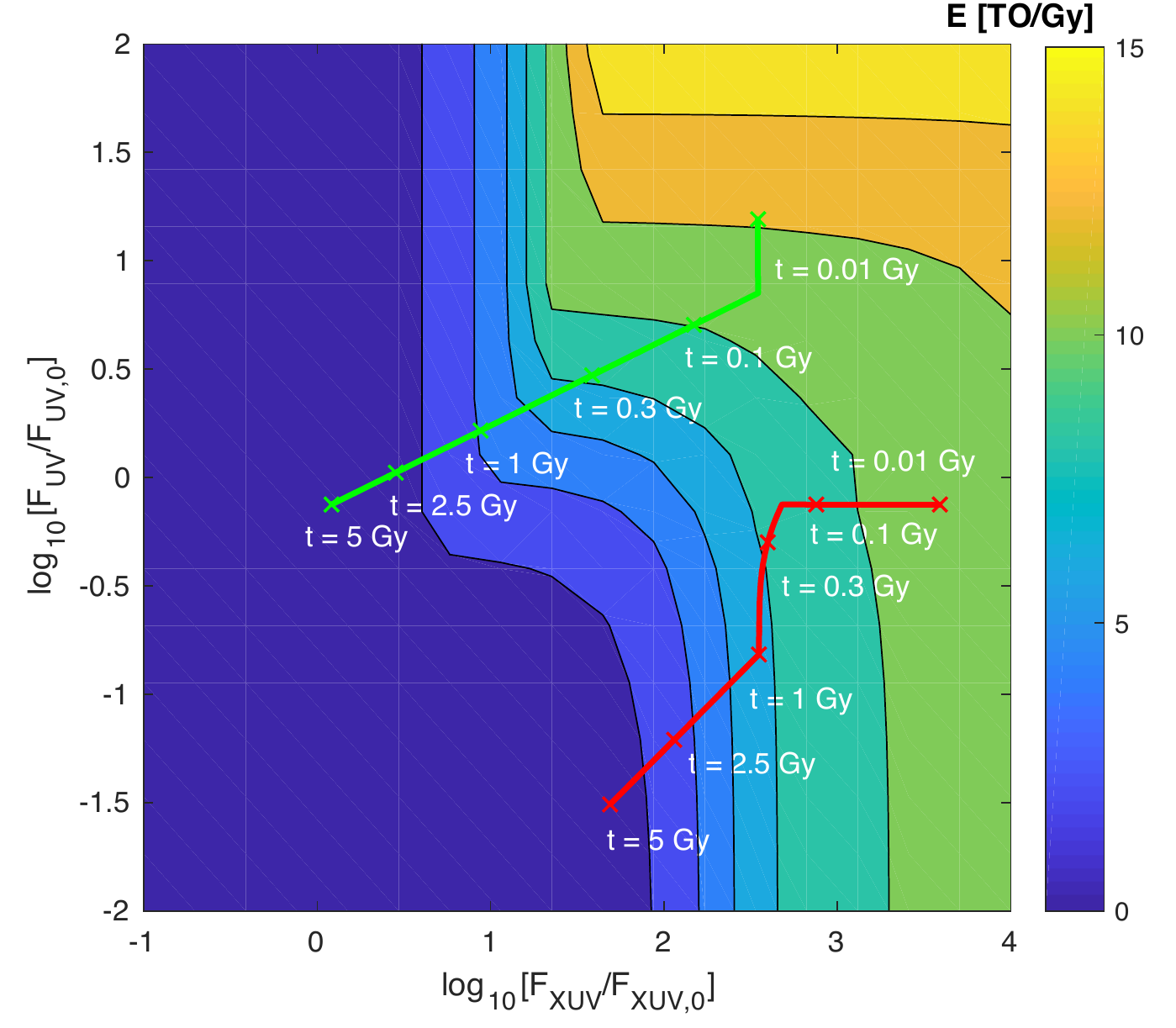}}
\else
		{\includegraphics[width=3.0in]{figures/XUV_UV_tracks_v6.pdf}}
\fi
	\end{center}
	\caption{Tracks showing modelled stellar XUV and UV evolution vs. time. The green line shows the Sun, while the red line shows Proxima Centauri for XUV model A. The Proxima data is scaled to represent a planet receiving a bolometric flux of 1366~W/m$^2$ at 5~Gy. For context, the contour plot of H escape rate as a function of stellar XUV and UV (Fig.~\ref{fig:phot_eff}e) is shown in the background.}
\label{fig:XUV_FUV_tracks}
\end{figure}

We now summarize the results of this section by calculating the atmospheric oxidation that would occur over a planet's history if the rate of exchange with the surface was zero ($k_1=k_2=0$) and \ce{H2O} was always abundant in the planet's upper atmosphere. For Venus, Earth and Mars, as well as a range of recently discovered low-mass exoplanets, we use equations (\ref{eq:Elim}-\ref{eq:esc_eff_3}) to calculate the integrated change in oxidation state of the volatile layer
\begin{equation}
N_a(t_1) = \int_{t_0}^{t_1} \frac{dN_a}{dt}dt= \int_{t_0}^{t_1} E dt.
\end{equation}
The integration is performed  starting from 10~My after the host star's formation, and assuming $N_a(t_0) = 0$. To derive upper and approximate lower limits on $E$ we take $t_1 = 5$~Gy and $t_1 = 100$~My, respectively. We also incorporate limits on escape due to both the diffusion rate and the escape efficiency. The dependence of $E$ on \ce{O2} build-up is neglected for now (this assumption is relaxed in the next section). We model the changing stellar luminosity $L$ using the data of \cite{Baraffe2015} and make similar assumptions on XUV evolution as in \cite{Schaefer2016}. Specifically, we assume an upper limit for XUV (model A) where $L_{XUV}=10^{-3}L$ for a set interval $t_{sat}$, then follows the power law $L_{XUV}=10^{-3}(t/t_{sat})^{-1.23}L$ \citep{Ribas2005} thereafter. 
We set $t_{sat}=50$~My for the Sun 
and $t_{sat} = 1$~Gy for M-class stars. We assume a lower limit (model B) where $L_{XUV}=10^{-3}L(t_1)$, then drops immediately to zero afterwards. Fig.~\ref{fig:XUV_FUV_tracks} shows G- and M-star evolutionary tracks vs. $F_{XUV}$ and $F_{UV}$, with the same plot of $E$  as in Fig.~\ref{fig:phot_eff}e) also shown for reference.

To derive upper and lower limits on oxidation, we combine XUV model A with escape efficiency $\epsilon_1$ and XUV model B with escape efficiency $\epsilon_2$, as described in the last subsection. For the TRAPPIST-1 planets, as a lower limit on XUV we use a constant $L_{XUV}=1.2\times 10^{20}$~W, 
based on recent XMM-Newton X-ray observations of the host star \citep{Wheatley2017}. 
Note that this XUV flux is considerably higher than the value 
assumed in \cite{Bolmont2016}, which was published before direct X-ray observations of the star were available. All data on exoplanet mass, radius and orbit and their host star luminosities is taken from the relevant discovery papers \citep{BertaThompson2015,Gillon2017,Dittmann2017}.

For the UV part of the spectrum, for the Sun we scale the spectrum based on the parametrization of \cite{Claire2012}. For the M-star exoplanets, we scale the XUV and UV portions using the synthetic spectrum for GJ832 from the MUSCLES database \citep{Loyd2016}, which is designed to be a proxy for Proxima Centauri. To incorporate UV time evolution, we incorporate the empirical time dependence formulae proposed by \cite{Shkolnik2014}, with a saturation point at 200~My age at 30 times the baseline UV value. For simplicity, we also use this formulation for the TRAPPIST planets, although we note that the UV evolution of very low mass M stars is still extremely uncertain. We incorporate the results in escape from the previous two subsections by assuming that once the XUV energy-limited escape rate (\ref{eq:Elim}) becomes greater than the diffusion limit $\Phi_{\ce H,diff}$, the latter sets the total H escape rate. 

\begin{figure}[h]
	\begin{center}
\ifarxiv
		{\includegraphics[width=3.5in]{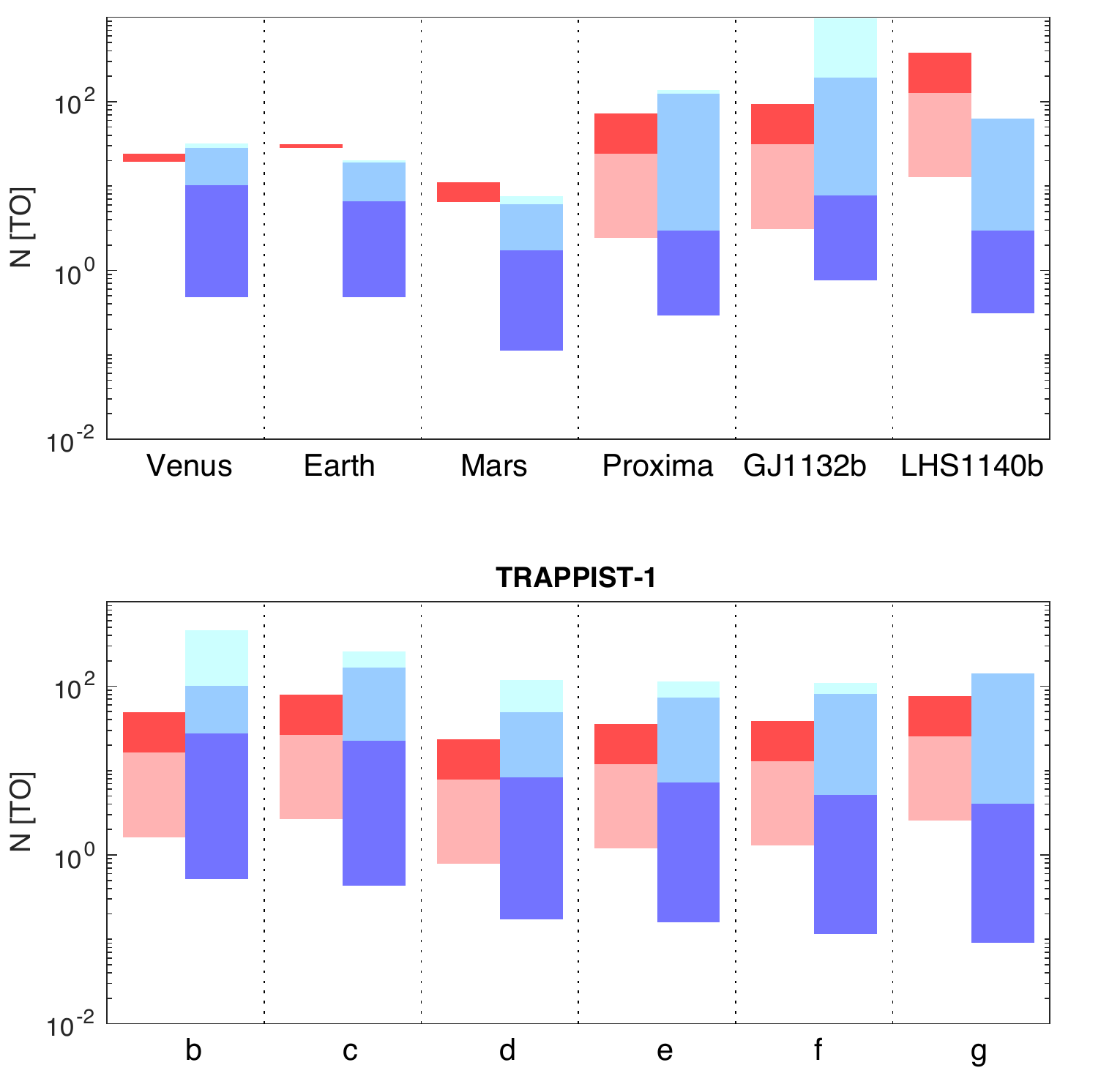}}
\else
		{\includegraphics[width=3.5in]{figures/bar_graph_v9.pdf}}
\fi
	\end{center}
	\caption{(top) Comparison between total potential oxidation of the volatile layer (blue) and reducing power (initial FeO content) of the silicate layer (red) for Venus, Earth, Mars, Proxima, Gliese 1132b and LHS1140b. (bottom) Same, but for the TRAPPIST-1b to g planets. Dark blue show the potential oxidation from H loss assuming diffusion and XUV energy limited H escape vs. time as in Fig.~\ref{fig:XUV_FUV_tracks}, with upper and lower bounds corresponding to $t_1=5$~Gy and 100~My, respectively. The peak of the medium blue gives an upper limit on oxidation, with $t_1=5$~Gy, $\epsilon = 0.3$, and O escape modelled according to \eqref{eq:num_flux_2}. The peak of the light blue shows the total amount of \ce{H2O} lost under the same upper limit assumptions. 
	Dark and light red show the standard and extreme limits on mantle FeO, respectively. Mantle mass fractions and mantle FeO content ranges for Earth, Venus and Mars are taken from \cite{Mcdonough1995}, \cite{Righter1996}, \cite{Javoy1999} and \cite{Robinson2001}. }
\label{fig:bar_graph}
\end{figure}

In Figure~\ref{fig:bar_graph}, limits on the potential oxidation $N_a(t_1)$ from hydrogen escape are shown in blue alongside estimates of the mantle reducing potential $|N_{b}(t_0)|$ in red, which we take to be equivalent to the initial FeO content here. Light blue shows the extreme upper limit on the total water (\ce{H2O}) lost {via H escape to space}, which is larger than the upper limit on $N_a(t_1)$ in situations where escape of oxygen with the escaping H is significant. On the exoplanet plots, a mantle FeO content range of 5-15~wt\% is assumed for the dark red bars, and the mantle mass fraction is set to 0.7. The light red denotes an `extreme lower limit' mantle FeO content of 0.5~wt\%, corresponding to values seen in some silicates on Mercury \citep{Zolotov2013}. As can be seen, the variation among the cases is significant, with $N_a(t_1)$ estimates varying from $\sim 0.1$ to $>100$ TO, while the $|N_b(t_0)|$ estimates range from 10s to 100s of TO.

The most striking aspect of Figure~\ref{fig:bar_graph} is that the potential mantle reducing power is comparable to, or much greater than, the potential oxidation $N_a(t_1)$ for many of the cases. It is also much greater than the conservative estimate of $N_a(t_1)$ (dark blue) in all cases. For all exoplanets except LHS1140b and TRAPPIST-1g, the extreme upper limit on total water loss is greater than $N_a(t_1)$. This is due to the fact that their high early XUV and relatively low UV keeps escape in the diffusion-limited regime (Fig.~\ref{fig:XUV_FUV_tracks}). All cases exhibit total potential water loss of more than 1~TO for $t_1=5$~Gy, although we stress that this assumes water \emph{never} becomes cold-trapped on the planet's surface. The huge reducing power of planetary mantles highlights the importance of performing coupled simulations of atmosphere--interior evolution, which we address next.

\section{Atmosphere-interior exchange: the magma ocean phase}\label{sec:melty}

To assess when XUV-driven H escape will actually drive $N_a$ to positive values and hence cause abiotic \ce{O2} buildup, we now relax our assumption that $k_{1,2}=0$ and turn to coupled atmosphere-interior simulations.  Rocky planet evolution can be divided into an early period, when the planet's silicate mantle is intermittently or permanently molten, and a much longer subsequent period once the mantle has solidified. Here, we use a similar approach to modelling the early magma ocean phase as in \cite{Schaefer2016}, with a few important modifications. 

First, we determine the interior structure of the planet as a function of its mass. Following \cite{Zeng2013}, we solve equations for interior radius $r$ and  pressure $p$ vs. mass $m$
\begin{equation}
\frac{dr}{dm}=\frac{1}{4\pi \rho r^2}
\end{equation}
\begin{equation}
\frac{dp}{dm}=-\frac{Gm}{4\pi r^4}.
\end{equation}
Here $G$ is the gravitational constant and $\rho$ is density. The equations are integrated from the core outwards until zero pressure is reached, and Newton's method is then used to find the correct core pressure for a given planetary mass. For the equation of state (EOS), we use a second-order Birch-Murnaghan equation with mantle and core coefficients determined from Earth data \citep{Dziewonski1981,Zeng2016}. This equation reproduces the mass-radius relationships of Earth, Venus and several low-mass exoplanets within observational error. It requires a value for the core mass fraction $f_c$, which we take to be 0.3 here. Sensitivity tests indicate low dependence of the results on the value of $f_c$ over a range of tens of percent.

By neglecting the dependence of interior pressure on temperature [$p=p(\rho)$ only], we greatly simplify our evolution calculations. The mantle temperature profile is calculated from the surface downwards as 
\begin{equation}
\frac{\partial T}{\partial p} = \frac{\alpha T}{\rho c_{p,m}}
\end{equation}
where $\alpha$, the thermal expansivity, is determined as in \cite{Abe1997} and $c_{p,m}$, the mantle specific heat capacity, is taken to be 1000~J/kg/K. 
Our approach neglects moist adiabat effects \citep{Abe1997}, which leads us to slightly underestimate the mantle melt fraction and hence overestimate magma ocean phase atmospheric \ce{O2} buildup. 

We calculate the local melt fraction by mass in the interior as
\begin{eqnarray}
\psi(r) = \left\{
  \begin{array}{lll}
   0 		 				  \quad&:&\quad  T  \le T_{sol}  \\
   \frac{T-T_{sol}}{T_{liq}-T_{sol}}  \quad&:&\quad T_{sol} <   T  < T_{liq}   \\
   1  						   \quad&:&\quad  T \ge T_{liq} .\\
  \end{array} \right.
\label{eq:melt_fraction}
\end{eqnarray}
where $T_{sol}$ and $T_{liq}$ are the solidus and liquidus temperature, respectively, which we determine using an extrapolation of the data of \cite{Hirschmann2000} as in \cite{Schaefer2016}. In real melts, the variation of $\psi$ with temperature is not as simple as represented by \eqref{eq:melt_fraction}, but the difference is not significant for our purposes \citep[for an insightful discussion of this issue, see][]{Miller1991}. We also make the standard assumption that the magma transitions to a high viscosity mush at a critical $\psi$ value, which we take to be $0.4$ here, and assume that the mush has no further contact with the liquid magma for the purposes of chemical equilibration \citep{Lebrun2013}.

We calculate the \emph{total} silicate layer melt fraction $\Psi$ as a function of surface temperature $T_s$ by numerically integrating (\ref{eq:melt_fraction}) in $r$ from the core-mantle boundary to the surface and normalizing to get 
\begin{equation}
\Psi(T_s) =  \frac{\int_{m(r_c)}^{m(r_p)}\psi(r)dm}{(1-f_c)M_p}. \label{eq:global_melt_fractionA}
\end{equation}
In our numerical model, $\Psi(T_s)$ is pre-calculated on a grid of $T_s$ values for a given planet mass, and the result at any $T_s$ is obtained when needed by interpolation. To perform an analytical check on our results (see Appendix~B), we have also parametrized it as
\begin{equation}
\Psi =  \frac 12 \left(\mbox{erf }\left[ \frac{T_s-T_t}{\Delta T}\right]+1\right) \label{eq:global_melt_fraction}
\end{equation}
where the parameters $T_t$ and $\Delta T$ are determined according to a least-squares fit. 

\begin{figure}[h]
	\begin{center}
\ifarxiv
		{\includegraphics[width=3.25in]{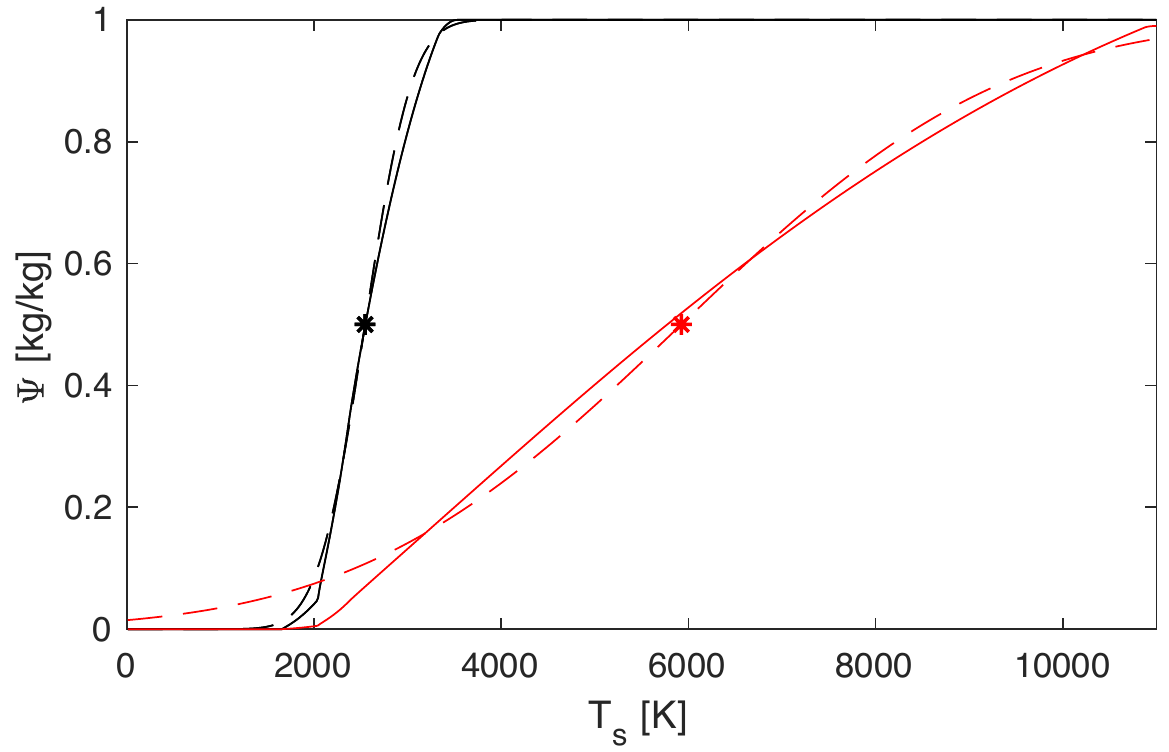}}
\else
		{\includegraphics[width=3.25in]{figures/melt_fraction.pdf}}
\fi
	\end{center}
	\caption{Plot of global mantle melt fraction as a function of surface temperature, for a 1~$M_E$ planet (black) and a 10~$M_E$ super-Earth (red). Solid and dashed lines show the numerical integration and the least squares fit of the results according to (\ref{eq:global_melt_fraction}), respectively. Asterisks show the surface temperature $T_t$ at which the global melt fraction is equal to 1/2 the mantle mass.}
\label{fig:melt_fraction}
\end{figure}

The total melt fraction $\Psi$ is shown as a function of surface temperature for a 1~$M_E$ and 10~$M_E$ planet in Figure~\ref{fig:melt_fraction}. At very high surface temperature, the base of the magma ocean is deep and the error associated with our extrapolation of the \cite{Hirschmann2000} solidus data is likely significant. However, the lower $T_s$ range where $\Psi$ varies  rapidly is the most important to atmospheric redox evolution. The total melt fraction is much smaller at a given surface temperature for more massive planets because their internal pressure increases more rapidly with depth.

The atmospheric thermal blanketing (greenhouse effect) is calculated using our line-by-line (LBL) climate model \citep{Wordsworth2016b,Schaefer2016,Wordsworth2017a}. The 2010 HITEMP line list is used to calculate opacities as a function of pressure, temperature and wavenumber. In keeping with our aim of calculating approximate upper limits on planetary oxidation, we consider only weakly reducing atmospheres dominated by \ce{H2O} and \ce{CO2} here. Addition of a \ce{H2} envelope would prevent \ce{O2} buildup until all \ce{H2} was lost to space, decreasing the final amount of atmospheric \ce{O2} in all cases. 

Our calculation uses 8000 spectral points from 1~cm$^{-1}$ to 5 times the Wien peak for a given surface temperature. 100 layers are used in the vertical, and an isothermal stratosphere at 200~K is assumed. The temperature profile is calculated as a dry adiabat in the lower atmosphere and a moist adiabat, when appropriate, in the high atmosphere, following the approach of \cite{Wordsworth2013b}. Specific heat capacity is calculated based on a specific concentration weighted average, with temperature variation accounted for using data from \cite{CRC2000}. For \ce{H2O}, we use the MT-CKD continuum version 2.5.2, while for \ce{CO2}, we use the GBB approach of \cite{Wordsworth2010} \citep{Baranov2004,Gruszka1998}. The \ce{CO2} continuum is not critical to our results as it is masked by water vapor lines at most temperatures and pressures.  We then use the LBL model to produce a grid of outgoing longwave radiation (OLR) values as a function of surface temperature $T_s$ and surface pressure $p_s$. The nominal planetary albedo $A$ is set to 0.3, although we test the sensitivity of our results to this parameter. Given the uncertainties in cloud processes, we regard this as a better approach for now than performing detailed shortwave calculations.

We solve for the thermal state of the interior vs. time using the energy balance at the top of the atmosphere
\begin{equation}
OLR(T_s,p_s) = \frac{(1-A)L(t)}{4\pi d^2}. \label{eq:TOA_ebal}
\end{equation}
Here $L$ is the time-dependent stellar luminosity and $d$ is the planet's semimajor axis. This approach neglects thermal transients due to the latent and sensible heat of the melt, which is a simpler approach than was taken in \cite{Schaefer2016}. Because a key focus here is understanding the pre-main sequence magma ocean phase of exoplanets around M-stars, we neglect the additional heating provided by accretion and radioactive decay. Previous work has shown that long half-life elements such as U, Th and K do not alter the magma ocean duration significantly, while short half-life elements such as \ce{^{26}Al} will only be important if both the accretion time and the star's pre-main sequence phase are short \citep{Lebrun2013}. Stellar luminosity as a function of time is calculated from the \cite{Baraffe2015} stellar evolution model dataset.

The solubility of \ce{H2O} in silicate melts is high and must be taken into account in any magma ocean model. We relate the surface pressure $p_v$ of \ce{H2O} to the mass fraction $q_v$ of \ce{H2O} in the melt as 
\begin{equation}
\frac{p_v}{p_{ref}} = \left(\frac{q_v}{q_{ref}}\right)^{1/\beta}.\label{eq:H2Omagsolu}
\end{equation}
Here $p_{ref}=24.15$~MPa, $q_{ref} = 0.01$~kg/kg and $\beta=0.74$ \citep{Papale1997,Schaefer2016}. 
Assuming that the amount of \ce{H2O} that becomes trapped in the solid mantle is small, the total \ce{H2O} mass equals the mass in the atmosphere $M_a$ plus that in the melt $M_v$
\begin{equation}
M_{tot} = M_a + M_v.
\end{equation}
Noting that $M_{tot}=q_{\ce{H2O}}M_{p}$, $M_a = 4\pi r_p^2 p_v / g$ and using (\ref{eq:H2Omagsolu}), the mass balance between atmosphere and magma ocean can be calculated by solving
\begin{equation}
0 = q_{\ce{H2O}}M_{p} - 4\pi r_p^2 p_v / g -  M_p(1-f_c)\Psi q_{ref}\left(\frac{p_v}{p_{ref}}\right)^{\beta}.\label{eq:mass_bal_MO}
\end{equation}
as a function of $p_v$. Equation (\ref{eq:mass_bal_MO}) implies that for an Earth-mass planet with an entirely molten mantle containing 10~TO of \ce{H2O}, the total atmospheric \ce{H2O} inventory will only be 0.2~TO.

For a given planet, we calculate atmospheric oxidation as a function of the starting \ce{H2O} and mantle \ce{FeO} inventories. We calculate the XUV-driven loss of \ce{H2O} vs. time using equations (\ref{eq:Elim}-\ref{eq:num_flux_3}) from Section~\ref{sec:escape}. We account for the effect of \ce{O2} buildup on H escape by assuming a simple linear dependence of $\Phi_{\ce H,diff}$ on $x_{\ce{O2}}$, such that 
\begin{equation}
\Phi_{\ce H,diff}=\left.\Phi_{\ce H,diff} \right|_{x_{\ce{O2}}=0}(1-x_{\ce{O2}}). 
\end{equation}
{This differs from our representation of the effects of \ce{O2} on H escape in \cite{Schaefer2016}, which was based on an analytic formula for diffusion of H in O \citep{Tian2015b}.} 
We solve for $T_s$ vs. time using a nested root-finding algorithm on (\ref{eq:TOA_ebal}) and  (\ref{eq:mass_bal_MO}) simultaneously. Finally, redox evolution is calculated by noting that because \ce{SiO2}, \ce{MgO} and \ce{Fe2O3} have net oxidizing power of zero in our classification scheme (Table~\ref{tab:redox_defs}), the total oxidizing power of the mantle is simply $N_{b}=\sum_i N_{b,i}p_i = - N_{\ce{FeO}}$, which is always less than or equal to zero. The rate of change of oxidizing power in the liquid part of the silicate layer (i.e. the magma ocean) is equal to twice the downward flow of liberated oxygen ($p_{\ce{O}}=+2$; see Table~\ref{tab:redox_defs}) from the atmosphere, minus the rate at which \ce{FeO} is lost due to mantle solidification
\begin{equation}
\frac{d N_{b,l}}{d t} = 2\Phi_{\downarrow,2} - N_{b,l} \frac{d \Psi}{d t}
\end{equation}

The mixing rates $k_1$ and $k_2$ between the magma ocean and atmosphere are assumed to be much smaller than a single model timestep. Conversely, we assume no mixing between the magma ocean and the high-viscosity part of the silicate layer (mush + solid with $\psi>0.4$). Because magma ocean crystallization begins at depth, the result for a planet that is steadily losing hydrogen is a mantle that becomes more oxidizing at larger radii \citep[see e.g. Fig.~10 in][]{Schaefer2016}. {Based on our mantle oxygen fugacity analysis (Section~\ref{sec:crusty}), we assume that \ce{O2} begins to accumulate in the atmosphere once the \ce{Fe^{3+}/(Fe^{2+} + Fe^{3+})} ratio in the magma ocean reaches 0.3 (see Fig.~\ref{fig:outgas_redox})}. Our evolution model is run until the pre-main sequence water loss phase finishes, which we assume occurs once the planet's absorbed stellar radiation (ASR) drops below the runaway greenhouse limit determined from the LBL climate data (around $ 282$~W/m$^2$ for an Earth-mass planet). 

\begin{figure*}[h]
	\begin{center}
\ifarxiv
		{\includegraphics[width=6.5in]{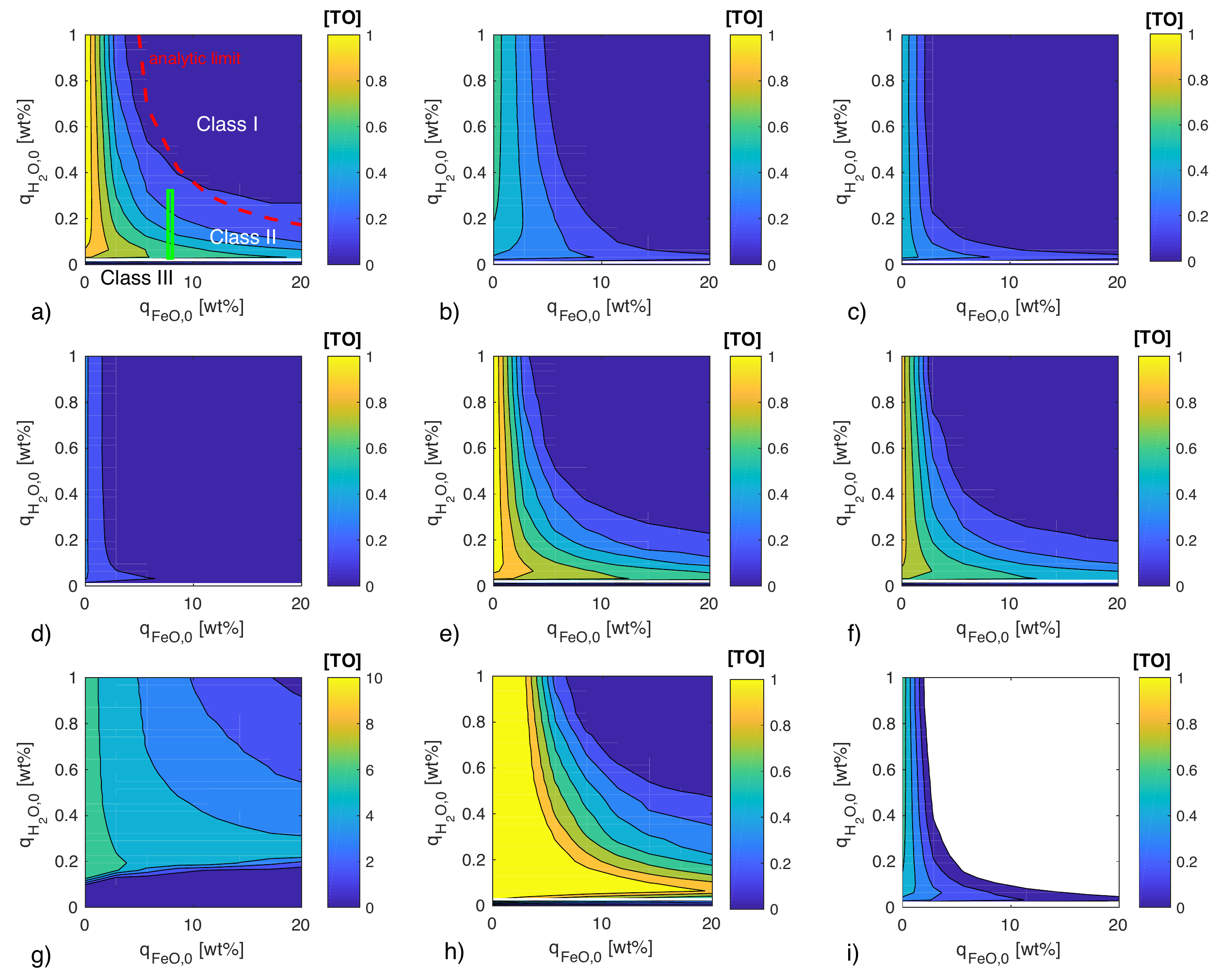}}
\else
		{\includegraphics[width=6.5in]{figures/PMS_O2_buildup_v6.pdf}}
\fi
	\end{center}
	\caption{a-g) Contour plots of atmospheric \ce{O2} immediately following the pre-main sequence runaway greenhouse phase of a planet orbiting an M-dwarf. The $x$-axis is initial specific concentration (weight fraction) of \ce{FeO} in the mantle, while the $y$-axis is the initial volatile layer \ce{H2O} inventory, expressed as a specific concentration relative to the entire planet mass. In a), the green box bounds the estimated mantle iron and \ce{H2O} abundance of Earth, while the red dashed line shows the analytic limit of \ce{O2} buildup given by equation~(\ref{eq:anal_soln_b}). Figs. b-e) are the same as a) except for b)  a 10~$M_E$ super-Earth, c) a planet receiving the same final stellar flux as Prox Cen b, d) a planet with albedo of  0.7 and e) a planet with 100~bars \ce{CO2} also present in the atmosphere. The contours in b) are scaled by a factor of 10 to account for the elevated planetary mass. f) and g) show escape rates as in a) except assuming O escape also occurs, at a rate determined by the photochemical model (f) or the direct application of the diffusion equations \eqref{eq:num_flux_1}-\eqref{eq:num_flux_2} with fixed $x_{\ce{O}}=1/3$ and $x_{\ce{O}}=2/3$ (g). Finally, (h) is the continuation of (a) for 5~Gy after the pre-main sequence phase, assuming a loss rate of oxidizing power to the interior of 0.2~TO/Gy.}
\label{fig:PMS_O2_buildup}
\end{figure*}

Figure~\ref{fig:PMS_O2_buildup} shows the model results for a range of cases as a function of the starting mantle \ce{FeO} mass fraction and global \ce{H2O} mass fraction. The colored contours show the atmospheric \ce{O2} buildup (equivalently, the value of $N_a$) in terrestrial ocean equivalent (TO) units. The red dashed line shows the analytic limit calculated according to (\ref{eq:anal_soln_b}) in Appendix~B. The match with the numerical model is not exact, but is close enough to demonstrate that we can correctly reproduce the essence of the model behaviour. 

The class labels describe the final volatile layer inventories and correspond to I: pure \ce{H2O}, II: \ce{O2} + \ce{H2O} and III: pure \ce{O2}. Class I planets begin with so much \ce{H2O} that they have molten surfaces until the very end of the pre-main sequence phase, and sufficient mantle FeO that all liberated O from the atmosphere is absorbed. Class II planets oxidize their upper mantles but retain some \ce{H2O}, leaving them with mixed \ce{O2}-\ce{H2O} atmospheres. Class III planets lose all of their water to space. As can be seen, in most of the plots, Class I, where little or no \ce{O2} is present, is the dominant evolutionary outcome. This regime may be the appropriate one for many of the TRAPPIST planets, based on recent analysis suggesting they have a water-rich interior composition \citep{Unterborn2018}.

Comparison of Figs.~\ref{fig:PMS_O2_buildup}a), c) and d) shows that pre-main sequence \ce{O2} buildup is most sensitive to a planet's orbit and its albedo. The reason for this is that both parameters strongly affect $t_{RG}$, the time at which stellar luminosity decreases enough for the planet to exit the runaway greenhouse phase. An increase in the planet's mass [Fig.~\ref{fig:PMS_O2_buildup}b)] leads to higher peak values of atmospheric \ce{O2} because diffusive separation of O and H is more effective under the higher gravity and the total mantle melt fraction is lower for a given surface temperature. However, the peak values normalized to the planet's mass are lower. The presence of moderate amounts of atmospheric \ce{CO2} (Fig.~\ref{fig:PMS_O2_buildup}e) has little effect on the results, in contrast to the situation for planets that are no longer in a runaway greenhouse state \citep{Wordsworth2013c}. Figs.~\ref{fig:PMS_O2_buildup}f) and g) show results as in (a) but allowing O escape according to diffusion limits from the photochemical model (f) and according to (\ref{eq:num_flux_1}) and (\ref{eq:num_flux_2}) with fixed $x_{\ce O} = 1/3$ and  $x_{\ce H} = 2/3$ {(g)}. The latter case shows \ce{O2} buildup several times greater than when the results of the 1D photochemical model are incorporated.

{Another effect we considered was possible differences in atmospheric chemistry due to the presence of species other than O and H. For example, in atmospheres where \ce{CO2} is present it will also photolyze, and  \ce{CO} and O will be produced as a result \citep{Yung1999}. Reactions such as \ce{CO + OH \to CO2 + H} could then enhance the stripping of hydrogen from \ce{H2O} and hence the H escape rate. We think that processes such as these are unlikely to dominate in the steam-dominated atmospheres we are considering here. However, they could plausibly alter H escape rates by some amount. While we leave detailed analysis of C-H-N-O atmospheres to future work, we can test the sensitivity of our results to such processes by altering the H escape rate by a fixed amount. Fig.~\ref{fig:PMS_O2_buildup}h) shows the result of such a simulation, where the H escape rate was increased by a factor of two in the diffusion-limited regime. As can be seen, the range of conditions under which \ce{O2} buildup is $>1$~TO increases, although for high starting \ce{H2O} and \ce{FeO} inventories buildup is still limited. This indicates that while additional study of the photochemistry in more complex systems is probably warranted, such effects are unlikely to change our basic conclusions.}

\begin{figure}[h]
	\begin{center}
\ifarxiv
		{\includegraphics[width=3.4in]{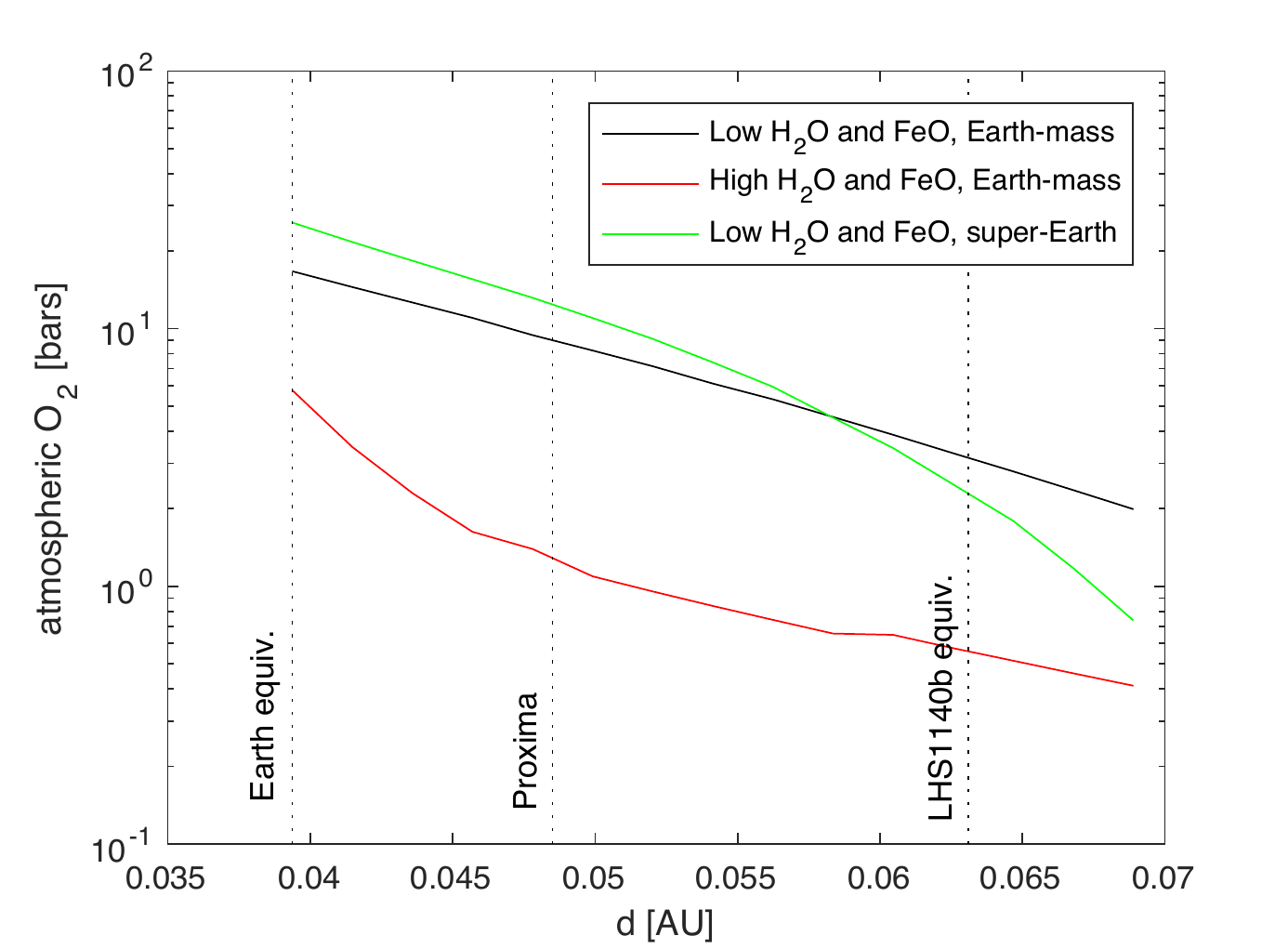}}
\else
		{\includegraphics[width=3.4in]{figures/PMS_O2_buildup_vs_dAU_v2.pdf}}
\fi
	\end{center}
	\caption{Plots of atmospheric \ce{O2} as a function of orbital distance, immediately following the pre-main sequence runaway greenhouse phase of a planet orbiting an M-dwarf. Colors correspond to (black) an Earth-mass planet with starting mantle inventory of $q_{\ce{FeO}} = 0.05$~kg/kg and $q_{\ce{H2O}} = 230$~ppmw (1~TO), (red) an Earth-mass planet with starting inventory of $q_{\ce{FeO}} = 0.1$~kg/kg and $q_{\ce{H2O}} = 2300$~ppmw, and (green) a 10$\times M_E$ super-Earth with starting inventory of $q_{\ce{FeO}} = 0.05$~kg/kg and $q_{\ce{H2O}} = 230$~ppmw (1~TO). }
\label{fig:PMS_O2_buildup_vs_dAU}
\end{figure}

{Figure~\ref{fig:PMS_O2_buildup_vs_dAU} shows the results of similar calculations to those in Fig.~\ref{fig:PMS_O2_buildup} as a function of received stellar flux, for a Proxima-like host star. This time, \ce{O2} buildup is expressed in terms of the resulting equivalent pressure in the atmosphere, in bars. The three lines show results for different planet masses and starting \ce{H2O}/\ce{FeO} mantle inventories. Note that in the high \ce{H2O} + \ce{FeO} super-Earth case, no \ce{O2} built up in the atmosphere at any of the orbital distances studied. The orbital distances required for the planet to receive Earth, Proxima b and LHS1140b equivalent fluxes are shown by the dotted lines. Clearly, \ce{O2} buildup is a very strong function of planet orbital distance, with more distant planets far less likely to develop thick \ce{O2} atmospheres. The two key reasons for this are that more distant planets a) receive fewer XUV photons and b) have shorter pre-main sequence runaway greenhouse phases.  The implications of this for future observations is discussed in Section~\ref{sec:diss}.}

Our results can also be compared with those of \cite{Luger2015}. That study, which was the first to highlight the importance of the M-star pre-main sequence phase to exoplanet abiotic \ce{O2}, predicted high \ce{O2} buildup but did not incorporate interaction between the atmosphere and interior. {For most parameter values, we find significantly lower atmospheric \ce{O2} abundances than were found in \cite{Luger2015}, demonstrating the importance of atmosphere-interior interactions}. It is also important to note that Figs.~\ref{fig:PMS_O2_buildup}a-{h}) capture an exoplanet's atmospheric state \emph{immediately after the magma ocean phase has finished}. All the exoplanets listed in Figure~\ref{fig:bar_graph} are likely several billion years old at least. If H escape ceased immediately after their initial runaway greenhouse phases finished, they would evolve to a very different atmospheric state subsequently. The post-runaway phase is studied in more detail next. However, as an example, Fig.~\ref{fig:PMS_O2_buildup}{i}) shows the atmospheric \ce{O2} from a) after 5~Gy has passed, assuming no further H escape (due to e.g. an effective \ce{N2}/\ce{CO2} cold trap) and a constant 0.2~TO/Gy loss rate of oxidizing power to the interior. Under these circumstances, the range of cases that continue to have residual atmospheric \ce{O2} from the magma ocean phase becomes very low. 

\section{Interior-atmosphere exchange after mantle solidification}\label{sec:crusty}

After a planet cools sufficiently to exit the runaway greenhouse state and the magma ocean freezes, a solid crust forms and mixing rates between the volatile and silicate layers decrease by orders of magnitude. Once this happens, any subsequent water loss may lead to atmosphere/ocean oxidation if the rate of redox exchange with the mantle is sufficiently low, even if the total reducing power of the mantle remains high. This period is hence particularly important to the question of abiotic \ce{O2} buildup. Many previous studies have analyzed redox exchange between atmospheric, oceanic, crustal and mantle reservoirs on Earth and Earth-like planets in some detail \citep[e.g., ][]{Holland2006,Zahnle2014,Domagal2014}, particularly in the context of the rise of oxygen on Earth \citep{Holland2006,Laakso2014,Laakso2017}. In keeping with the overall approach of this paper, here we constrain the redox budget for a wide range of planetary conditions in a simple way, rather than performing detailed modeling of Earth-specific processes. 

Based on the definitions in Table~\ref{tab:redox_defs}, atmospheric \ce{O2} buildup will commence once the volatile layer oxidizing power $N_a$ becomes positive\footnote{Note that our definition of the volatile layer includes both the atmosphere and a liquid \ce{H2O} ocean, when present. However \ce{O2} is relatively insoluble in water, with around 70~TO required on Earth to dissolve 50\% of Earth's present-day atmospheric \ce{O2} content \citep{Luger2015}. 
Hence we treat buildup of \ce{O2} in the volatile layer and the atmosphere as equivalent here.}. From \eqref{eq:basic1}, an increasing trend in $N_a$ corresponds to $E - k_1 N_{a} + k_2 N_{b} > 0$. This will occur if oxidation via atmospheric loss of hydrogen $E$ outpaces subduction of oxidized crust and outgassing from a reducing mantle, or if the mantle itself is so oxidized that it can directly outgas \ce{O2}.

The outgassing term is proportional to the total rate of volcanism $k_2$ and to the mantle redox state $N_b$. In general both terms will be spatially heterogeneous, but we omit this complication here. Assuming a redox budget dominated by H, O and Fe, a constraint on the \ce{H2O} outgassing rate allows the oxidizing power of volcanic gases to be estimated as a function of mantle redox state, based the equilibrium 
\begin{equation}
\ce{H2O <=> H2 + \frac12 O2 }. \label{eq:H2Oequi}
\end{equation}
Given an equilibrium constant $K_{eq}$ for \eqref{eq:H2Oequi}, we can write an expression for hydrogen molar concentration
\begin{equation}
x_{\ce{H2}} = \frac{p_{\ce{H2}}}{p_{\ce{H2}}+p_{\ce{H2O}}+p_{\ce{O2}}}\approx \frac{R}{R+1 }; \qquad R = \sqrt{\frac{K_{eq}}{f_{\ce{O2}}}}.\label{eq:H2Oequi_bal}
\end{equation}
Here $f_{\ce{O2}}$ is the oxygen fugacity of the magma \citep{Lindsley1991}, which is the same as the partial pressure $p_{\ce{O2}}$ under ideal gas conditions. The second approximate equality in \eqref{eq:H2Oequi_bal} is true as long as $p_{\ce{O2}}<p_{\ce{H2O}}$. For typical outgassing temperature ($T=1450$~K), $K_{eq}=1.9 \times 10^{-7}$~Pa\footnote{Our calculation here roughly follows the approach taken in \cite{Ramirez2014a}. A more complete calculation would account for the pressure dependence of  $K_{eq}$. Analysis of JANAF data shows that $K_{eq}$ increases with pressure, leading to higher rates of \ce{H2} outgassing.}. 

\begin{figure}[h]
	\begin{center}
\ifarxiv
		{\includegraphics[width=3.5in]{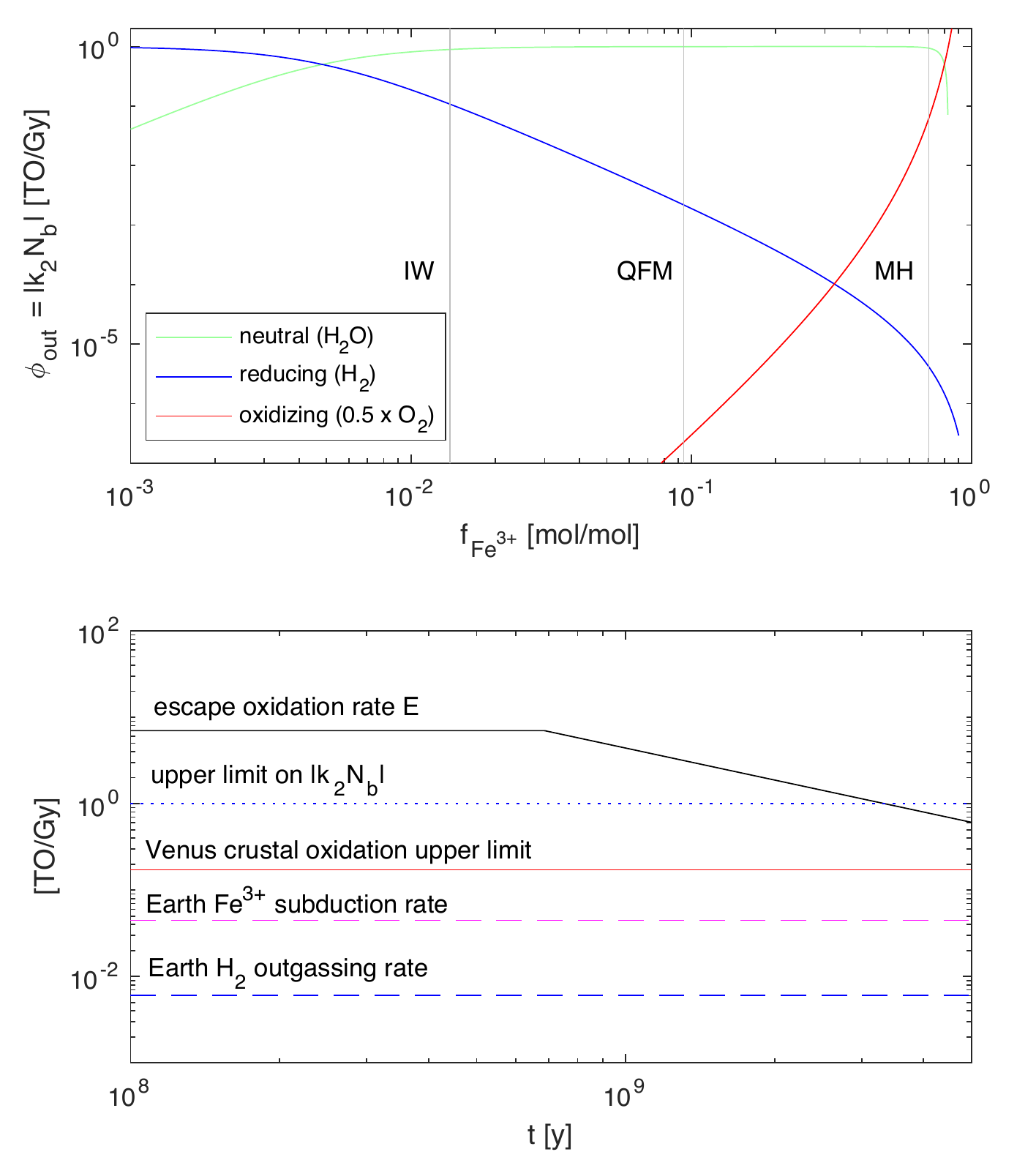}}
\else
		{\includegraphics[width=3.5in]{figures/outgassing_redox_v6.pdf}}
\fi
	\end{center}
	\caption{(top) Volatile layer redox changes due to volcanic outgassing vs. magma molar ratio of \ce{Fe^{3+}} to total \ce{Fe} for a planet with present-day Earth's outgassing rate. The labelled gray lines show several mineral redox buffers  (IW = iron-w\"ustite; QFM = quartz-fayalite-magnetite; MH = magnetite-hematite), based on the data in \cite{Lindsley1991}. (bottom) Volatile layer redox changes on Earth and Venus due to outgassing and subduction, compared with the H escape oxidation rate $E$ for Earth, assuming no cold-trapping. The Venus upper limit is obtained assuming a volcanism rate of 10~km$^3$/y, with 100\% of the magma oxidizing on contact with the atmosphere. }
\label{fig:outgas_redox}
\end{figure}

We relate oxygen fugacity $f_{\ce{O2}}$ to the iron oxidation ratio of the magma $x_{\ce{Fe^{3+}}}/x_{\ce{Fe_{tot}}}$ and temperature $T$ using the empirical formula from \cite{Zhang2017}. This formula is based on experimental data from mafic (metal-rich) silicate melts of the type expected for a wide range of volcanic scenarios\footnote{Previous work \citep[e.g., ][]{Kress1991} has shown that the redox state of volcanic gases depends to some extent on the abundance of additional compounds such as \ce{Al2O3} and \ce{MgO}. We ignore this extra source of complexity here.}. Figure~\ref{fig:outgas_redox} shows the results of this calculation. As can be seen, volcanic gases are reducing for $x_{\ce{Fe^{3+}}}/x_{\ce{Fe_{tot}}}$ values below around 0.3; i.e. all but the most oxidized magmas. Various mineral redox buffers are also displayed on the plot. The oxidation state of Earth's upper mantle is close to the quartz-fayalite-magnetite (QFM) buffer \citep{Frost2008}, while those of Venus and Mars' mantles are most likely around magnetite-hematite (MH) and iron-w\"ustite (IW), respectively \citep{Florensky1983,Fegley1997,Wadhwa2001,Wordsworth2016}.  Clearly, the final \ce{Fe^{3+}}/\ce{Fe} ratio of a planet's mantle after its magma ocean phase ceases is critical to its subsequent atmospheric evolution. Planets with \ce{Fe^{3+}}/\ce{Fe}$<5\times 10^{-3}$ primarily outgas \ce{H2} rather than \ce{H2O}, while those with \ce{Fe^{3+}}/\ce{Fe}$>$0.3 are so oxidizing that they can outgas \ce{O2} directly in significant amounts. {Clearly, even if a planet does not build up an \ce{O2} atmosphere from H loss during its magma ocean phase, oxidation of the upper mantle will decrease the reducing power of volcanic gases and hence can facilitate later buildup of an \ce{O2} atmosphere through other mechanisms}.

The total \emph{rate} of volcanism as a function of time on a rocky planet is  challenging to calculate from first principles. The extent to which exoplanets can be expected to exist in plate-tectonic, stagnant lid or other geodynamical regimes is still a subject of considerable controversy in the literature \citep[e.g,][]{Valencia2007,Korenaga2010,Weller2012}. Indeed, in some situations strong dependence on initial conditions and hysteresis effects are expected \citep{Weller2012}.  Given this, we regard it as wisest to use constraints from previous modeling and observations of Earth and Venus and do not attempt our own detailed modeling here.

Efficient volcanic outgassing requires a) a high rate of mantle melting and b) efficient degassing of the melt once it is close to the surface. Following magma ocean solidification, the primary controls on the mantle melting rate are the mantle temperature and water abundance. Immediately after magma ocean solidification, the mantle will be much hotter than on the present-day Earth. This could potentially lead to a rate of volcanism up to 10-50 times Earth's present-day rate during the first 2~Gyr \citep{Kite2009}. 
Assuming that this volcanism is associated with the same outgassing {rate and composition} as on Earth today \citep[$1.4 \times 10^{13}$ mol water/yr; ][]{Parai2012}, this would result in \ce{H2O} outgassing rates of 2-9~TO/yr.  
However, degassing of these melts is not assured. Stagnant-lid planets and planets with thick volatile layers (atmosphere or oceans) both suppress degassing from melts due to overburden pressure\footnote{Overburden pressure also has redox implications. For example, a planet with several times Earth's ocean inventory but the same mantle redox state would potentially outgas hydrogen at a significantly lower rate. This could lead to oxidation of the volatile layer via H escape over time even given quite modest rates of H escape. Further modeling is required to assess this possibility quantitatively.}. Furthermore, hotter mantles may develop more sluggish plate tectonics and thicker crusts due to dehydration of the mantle following melt formation \citep{Korenaga2003,Korenaga2017}. We therefore take $|k_2N_b|\approx 1$~TO/Gy as an upper limit on the possible outgassing rate, corresponding to a planet with vigorous plate tectonics and mantle oxygen fugacity around the iron-w\"ustite buffer.

For comparison, on present-day Earth (which has plate tectonics) an upper limit on the outgassing rate of \ce{H2O} can be taken as  $1.8\times 10^{13}$ to $1\times10^{14}$~mol/y, or 0.23--1.33~TO/Gy \citep{Jarrard2003,vanKeken2011}. 
 On Venus, which is currently in a stagnant or episodic lid regime, the production rate of crust averaged over geological time is likely around 5~km$^{3}$/y based on the atmospheric $^{40}$Ar abundance, with an upper limit rate of volcanism from modeling of around 10~km$^3$/y \citep{Gillmann2009}. All things equal, this translates to 0.2-2 times Earth's outgassing rate, 
 although Venus' overburden pressure from the 92~bar \ce{CO2} atmosphere likely inhibits \ce{H2O} release from magma \citep{Head1986}. A crustal oxidation upper limit can be determined using the crustal production rate and the estimated concentration of FeO in the Venusian mantle.

The final important term in the post-magma ocean phase is the rate of removal of oxidizing power from the volatile layer $k_1 N_a$. On planets whose redox balance is dominated by H, O and Fe, as we are assuming here, the rate of removal of oxidizing power from the atmosphere and crust is primarily determined by the flux of \ce{Fe^{3+}} to the mantle. This has been estimated for the present-day Earth as $12\times10^3$~kg/s, or $k_1N_a \approx 0.04$~TO/Gy  \citep{Lecuyer1999}. On the early Earth, mantle convection rates were probably higher, which could have resulted in somewhat higher values than this.

Figure~\ref{fig:outgas_redox} plots the limits we have just discussed vs. time alongside the maximum oxidation rate via H escape, $E$, estimated from Fig.~\ref{fig:phot_eff}. As can be seen, on many planets fractions of a TO or more of oxidizing power can be removed from the atmosphere via interior exchange. Hence in M-star systems older than a few Gy, only a few planets will retain atmospheric \ce{O2} produced during the pre-main sequence phase. Nonetheless, the continued hydrogen escape rate is significantly greater than the outgassing and subduction terms under most conditions. Hence many planets will build up abiotic atmospheric \ce{O2} later in their evolution unless they have an effective cold trap to keep \ce{H2O} (and other H-bearing gases) locked in the lower atmosphere.  We discuss the cold-trapping process further next.

\section{The key role of tropospheric cold-trapping}\label{sec:trap}

In Earth's present-day atmosphere, \ce{H2O} is cold-trapped, keeping the stratosphere dry and ensuring that the rate of oxidation is limited by H diffusion through the homopause, rather than the strength of incident UV or XUV stellar radiation. As has been previously demonstrated \citep{Wordsworth2013c,Wordsworth2014}, the partial pressure of non-condensing volatiles (primarily \ce{N2} and \ce{O2}) in Earth's atmosphere is critical to the efficiency of this cold trap. If escape of heavy gases is efficient on a given exoplanet, the cold trap will be removed and hydrogen loss can become rapid again.

A rough guide to the surface \ce{N2}/\ce{O2} partial pressure below which cold-trapping ceases to be effective is obtained by setting the moist convection number \citep{Wordsworth2013c}
\begin{equation}
\mathcal M = \epsilon p_v L \slash p_n c_p T_s\label{eq:Meqn}
\end{equation}
equal to unity. Here $L$ is the specific latent heat of the condensing gas, $c_p$ is the specific heat capacity at constant pressure of the non-condensing gas (or gas mixture), $T_s$ is temperature, $p_v$ and $p_n$ are respectively the partial pressures of the condensing and non-condensing gases in the atmosphere, $\epsilon = m_v\slash m_n$ is the molar mass ratio between the two gases, and all values are defined at the surface. For a potentially habitable planet with $T_s=290$~K, $\mathcal M = 1$ yields $p_n \approx 0.1$~bar, or around $ 1\slash 8^{\mbox{th}}$ of Earth's current atmospheric \ce{N2} inventory. 

Atmospheric escape of heavy gases from planets around M-stars has recently been studied by several groups. \cite{Dong2017} studied ion escape from Proxima Centauri b powered by stellar wind activity, and concluded that several bars of oxygen could have been lost over the planet's lifetime. In addition, \cite{Airapetian2017} modeled the XUV-driven non-thermal ion escape of oxygen and nitrogen and found that 10-100s of bars of these gases could be lost from Earth-like planets, provided that the upper atmosphere was hydrogen-poor. { \cite{Garcia2017} studied \ce{H^+} and \ce{O^+} escape using a slightly different  model and came to similar conclusions as \cite{Dong2017} regarding the overall loss rate}.  Further modeling to account for the complex radiative processes that can occur in ionized N- and O-rich atmospheres will be useful, but at this time it appears that the loss rates of \ce{N2} from habitable zone planets around M-stars may be rapid in many cases. {This means that H loss and abiotic \ce{O2} buildup could occur on many of these planets even if they avoid extreme oxidation during their magma ocean phases}. Clearly, methods to detect the partial pressure of \ce{N2} remotely on exoplanets \citep[e.g., ][]{Schwieterman2015} should be considered seriously in future mission planning.

\section{Abiotic oxidation due to atmospheric species other than \ce{H2O}}\label{sec:other_spec}

Cold-trapping is only effective if the volatility of the hydrogen host molecule is low for the given atmospheric thermal structure. For terrestrial-type planets with surface liquid water, this means that \ce{H2}, \ce{CH4}, \ce{H2S} and \ce{NH3} are all species that can cause planetary oxidation when emitted to the atmosphere in significant quantities. \ce{H2} emission from volcanic outgassing has already been considered in Section~\ref{sec:crusty}. Another possibility for \ce{H2} emission is crustal serpentinization. To occur, this process requires the presence of significant quantities of iron in the crust in an intermediate oxidation state (e.g. as the mineral fayalite, \ce{Fe2SiO4}) in direct contact with liquid water. Based on extrapolation of measurements of terrestrial ophiolites \citep[e.g., ][]{Etiope2013}, this could plausibly lead to loss of fractions of a terrestrial ocean's worth of \ce{H2O}, depending on the planet's crustal recycling rate.

Methane is another interesting case. The role of \emph{biogenic} methane in driving the irreversible oxidation of Earth during the Archean and Proterozoic eras has been studied previously \citep{Catling2001}. Based on coupled ecology-climate-chemistry models of the Archean, it has been estimated that \ce{CH4} levels could have built up to 1000~ppm; enough to drive the oxidation of Earth by up to 0.17~TO/Gy \citep{Catling2001,Kharecha2005}. However, on sterile planets, the steady state abundance of \ce{CH4} is likely to be orders of magnitude lower than this.
One further possibility is \ce{CH4} clathrate formation, which has been hypothesized as an explanation for Titan's atmospheric composition \citep{Tobie2006} and for past episodic warming on Mars \citep{Wordsworth2017a}. For episodically frozen waterworlds, clathration leads to particularly interesting possibilities. For example, a frozen planet with low outgassing rates and a minimal cold-trap might simultaneously build up atmospheric oxygen and subsurface \ce{CH4} clathrate deposits. If these deposits were later destabilized by external perturbations, transient atmospheres containing both \ce{O2} and \ce{CH4} would result. Because \ce{CH4} reacts quite rapidly with \ce{O2} on geological timescales, these cases are likely to be short-lived. Diffusion of oxidized gases  from the atmosphere into subsurface ice might also proceed more rapidly than hydrogen escape in many cases. Nonetheless, detailed modeling of this possibility in future would be interesting.

Finally, ammonia and hydrogen sulfide are present only in trace quantities on the present-day Earth and do not contribute significantly to hydrogen loss. Ammonia is easily photolyzed by ultraviolet light and is not likely to be abundant enough to cause significant hydrogen escape on either abiotic or inhabited planets with Earth-like ocean volumes. The extreme chemical stability of the \ce{N2} molecule \citep[e.g., ][]{Moses2013b,Wordsworth2016} means that scenarios where N acts as an effective shuttle to carry H from surface reservoirs past the cold trap are hard to sustain. Scenarios in which \ce{H2S} abundances build up to levels sufficient to drive gross oxidation also seem unlikely, based on upper-limit estimates of outgassing on S-rich planets such as Mars \citep[e.g., ][]{Halevy2014} and the relatively high solubility of sulfur species in \ce{H2O}, which generally leads to rapid atmospheric removal via rainout.

\section{Discussion}\label{sec:diss}

\begin{table*}
\begin{center}
\begin{tabular}{|c|c|c|}
\hline 
\textbf{Planet} & \textbf{Abiotic \ce{O2} buildup potential} & \textbf{Remarks} \\
\hline 
Prox Cen b & \color{blueish} \textbf{MEDIUM} & Low received stellar flux, Earth-like mass. \\
\hline 
GJ1132b & \color{red} \textbf{HIGH} & High stellar flux: planet is likely sterile. \\
\hline 
LHS1140b & \color{green} \textbf{LOW} & Low stellar flux, high planet mass. \\
\hline 
TRAPPIST-1b & \color{blueish} \textbf{MEDIUM} & High stellar flux: planet is likely sterile.   \\
\hline 
TRAPPIST-1c & \color{blueish} \textbf{MEDIUM} & High stellar flux. \\
\hline 
TRAPPIST-1d & \color{blueish} \textbf{MEDIUM} & Moderate stellar flux.   \\
\hline 
TRAPPIST-1e & \color{blueish} \textbf{MEDIUM} & Moderate stellar flux.   \\
\hline 
TRAPPIST-1f & \color{green} \textbf{LOW} & Low stellar flux.    \\
\hline 
TRAPPIST-1g & \color{green} \textbf{LOW} & Low stellar flux.    \\
\hline 
\end{tabular}
\end{center}
\caption{Qualitative summary of the implications of our results for a range of nearby low mass exoplanets. }
\label{tab:exec_summary}
\end{table*}

One of the key motivations of this study was to determine the situations in which atmospheric \ce{O2} can be regarded as a biosignature, i.e. a reliable indication of the presence of life. Our results indicate that while the presence of \ce{O2} alone should never be regarded as a `smoking gun', it is far more likely to have a biological origin in some cases than in others. Most importantly, planets that orbit further from their host stars will lose less hydrogen and \ce{N2} to space and are likely to have higher mantle FeO content \citep{Robinson2001,Fischer2017}, making them much less likely to build up long-lived abiotic \ce{O2} atmospheres. 

We find that the pre-main sequence phase of M-dwarfs can lead to significant water loss, in agreement with previous work \citep{Ramirez2014b,Luger2015,Tian2015}. Our photochemical calculations have shown that the rate of H reaching the upper atmosphere via diffusion is the main limit on oxidation when the stellar UV/XUV ratio is low or \ce{O2} has built up to high abundance in the atmosphere. However, escape can still be quite rapid even in \ce{O2}-rich atmospheres. The presence of a hydrogen corona around an exoplanet is therefore still compatible with an \ce{O2}-rich atmosphere.

Interaction of the atmosphere with the planet's interior is critical to understanding whether or not an abiotic \ce{O2} atmosphere will build up. In contrast to \cite{Luger2015}, we find that for a wide range of habitable zone planets, pre-main sequence water loss leads to little or no atmospheric \ce{O2} build-up. The oxygen liberated from \ce{H2O} photolysis instead mostly reacts with iron in the mantle, which is molten due to the strong greenhouse effect of the planet's steam atmosphere. However, once the planet has cooled and forms oceans and a crust, redox exchange rates between the atmosphere and interior decrease by orders of magnitude, and if the planet lacks a cold trap, hydrogen loss can still lead to abiotic \ce{O2} build-up in many cases \citep{Wordsworth2014}. The issue of how efficiently habitable zone planets lose `non-condensing' species such as \ce{N2} and \ce{CO2} to space or their interiors is therefore critical to planetary redox evolution, and future observational planning should emphasize ways to constrain the atmospheric abundance of these species. 

Planetary oxidation is vital not only to biosignature analysis, but also to the question of whether life can originate on a given planet in the first place. Hyper-oxidized planets are likely to be poor places for life to begin, as are planets that remain reducing enough to guard a significant hydrogen envelope. The period in which a planet's surface transitions from strongly reducing to oxidizing conditions is likely to be the ideal period for biogenesis  \citep{Wordsworth2012a} --- particularly if it involves {local} redox heterogeneity on the surface. The concept of a `goldilocks zone' for planetary redox is in our view as important as the better-studied habitable zone for liquid water \citep{Kasting1993}.

Table~\ref{tab:exec_summary} gives a qualitative summary of the implications of our results for a range of the lowest mass exoplanets in the nearest stellar systems currently known. Despite the many uncertainties, we have provided our overall estimate of the abiotic \ce{O2} buildup potential for each planet as a guide for future work. Of the two planets discovered by the MEarth team, GJ1132b is a good candidate for abiotic \ce{O2} build-up, as we have argued in \cite{Schaefer2016}, while LHS1140b is much less likely to develop an abiotic oxygen atmosphere due to its greater orbital distance and higher mass.  LHS1140b receives about the same flux as Mars, placing it in the nominal \ce{N2}-\ce{H2O}-\ce{CO2} habitable zone. Detection of an oxygen-rich atmosphere on this planet would hence be extremely exciting, as it would be unlikely to be due to abiotic processes alone.

For the TRAPPIST planets, the same trends of \ce{O2} buildup with orbital distance and planet mass apply as for the other cases. The low observed densities of many of the planets suggest they may have retained a significant volatile component, perhaps as a result of migration, although the uncertainties in these measurements mean no definite conclusions can be made at present. Retention of an \ce{H2} envelope clearly precludes the presence of \ce{O2}, while a thick \ce{H2O} layer would move the planets towards the `Class I' regime of Fig.~\ref{fig:PMS_O2_buildup}, where \ce{O2} buildup is also inhibited. Hence if the TRAPPIST planet densities are confirmed, abiotic oxygen buildup on them appears unlikely, which makes them (like LHS1140b) very interesting future targets for atmospheric characterization.


{In this paper we have placed special emphasis on the redox evolution of exoplanets around M-stars, because these are the cases for which observational tests to our model will come first. However, our results are also applicable to exoplanets around other star types. Because planets that orbit G-stars receive fewer XUV photons over their lifetimes for a given total stellar flux, they will undergo less total oxidation in the majority of cases. This is clear from Fig.~\ref{fig:bar_graph}, which shows lower total potential oxidation for Venus, Earth and Mars than for the nine exoplanets studied, all of which orbit M-class stars. As Fig.~\ref{fig:outgas_redox} shows, abiotic oxygen atmospheres still have the potential to build up on planets around G-stars, but the probability is lower for the majority of cases. Given the challenges to biogenesis and biosignature detection that extreme oxidation poses, this general difference between M- and G-star planets motivates the long-term development of missions to study Earth-like exoplanets around Sun-like stars.}

\section{Future work}\label{sec:future}

Improvements in our understanding of planetary redox evolution in future will require developments in several key areas. First, further detail in multi-species escape modeling is still required, including chemistry, conduction, diffusion and radiative cooling effects. The way in which 3D dynamics affects water loss in the cold-trap regime and the planetary radiation balance also needs further modeling. Regarding interior processes, further experimental constraints at high pressure and temperature are required to understand core-mantle equilibration and Fe redox disproportionation, and hence the initial oxidation state of a planet's mantle. Finally, we note that the generalized framework we have proposed in Section~\ref{sec:general} can and should be extended to other species, such as carbon and sulfur, in future.

Ultimately, the most powerful constraints on the modeling described here will come from direct observations. Information on atmospheric composition for hot, sterile planets inside the runaway greenhouse limit, such as TRAPPIST-1b and GJ1132b, will be particularly critical, as it will allow testing and calibration of our models in cases where biologically produced oxygen is not possible. The James Webb Space Telescope, which launches in late 2018, will have the ability to characterize the atmospheric composition of most of the planets we have modeled here via a combination of thermal emission and transmission spectroscopy \citep{Morley2017}. Combined with chemical modeling, such observations may allow the oxidation state of the atmosphere $N_a$ to be retrieved. Direct detection of atmospheric \ce{O2} itself will be possible via ground-based high dispersion spectroscopy starting in the early 2020s, or by future direct imaging missions \citep{Snellen2013,Rodler2014,Meadows2017}.

\acknowledgments
RW acknowledges funding from the Kavli foundation and the Star Family Challenge for Promising Scientific Research. RAF is funded by NASA grant NNX17AE27G and the Henry Luce Foundation. This article has benefitted from discussion with many researchers, including Dan Schrag and Sara Seager on abiotic oxygen, Bob Johnson on atmospheric escape, and Daniel Jacob and Eric H\'ebrard on photochemistry. {The authors also thank Feng Tian and an anonymous reviewer for providing useful critical feedback on an earlier version of the manuscript}.

\begin{appendices}

\section{Hydrodynamic escape of a binary gas mixture}\label{apx:diff}

Because clear derivations of the equations for hydrodynamic escape of a gas mixture from first principles are scarce in the literature, we present details of our own derivation here. We begin from the general equation for diffusion in a binary mixture
\begin{equation}
\mathbf u_1 - \mathbf u_2 = -\frac{1}{x_1x_2}\frac{b}n\left(  \mathbf d_{12}+ k_T\nabla \log T \right)\label{eq:start_diff_eq}
\end{equation}
where
\begin{equation}
 \mathbf d_{12} = \nabla x_1 + \frac{n_1n_2(m_2-m_1)}{n\rho}\nabla \log p - \frac{\rho_1\rho_2}{p\rho}(\mathbf F_1 - \mathbf F_2) .\label{eq:d12}
\end{equation}
Here $\mathbf u_i$, $x_i$, $n_i$, $\rho_i$ and $m_i$ are the relative velocity, molar concentration, number density, mass density and molecular mass of species $i$, respectively.  $n$, $\rho$, $p$ and $T$ are the total number density,  mass density, pressure and temperature, respectively. Finally,  $b$ is the binary diffusion coefficient between the two species, and $k_T$ is the thermal diffusion ratio. The derivation of \eqref{eq:start_diff_eq} from the Maxwell-Boltzmann equation is described in detail in \cite{Chapman1970}. 

The momentum equation for a binary mixture \citep[][ eq. 8.21, 4]{Chapman1970} is 
\begin{equation}
 \nabla p =   \rho_1 \mathbf F_1 + \rho_2 \mathbf F_2 - \rho \frac{D_0\mathbf u_0}{Dt} 
\end{equation}
where $\mathbf u_0$ is the mass-weighted mean velocity of the flow, and $D_0$ the corresponding advective operator. We can substitute this expression into \eqref{eq:d12} and rearrange to get 
\begin{equation}
 \mathbf d_{12} =\left(\nabla  p_1 - \rho_1 \mathbf F_1' \right)/ p . \label{eq:d12new}
\end{equation}
where we have defined the acceleration $\mathbf F_1'$ in the Lagrangian frame following the flow as 
\begin{equation}
 \mathbf F_1'  \equiv \mathbf F_1 -  \frac{D_0\mathbf u_0}{Dt}.\label{eq:F1dash}
\end{equation}
Note we have also used the definition of partial pressure $p_1=x_1p$ and total density $\rho=\rho_1+\rho_2$. Substituting \eqref{eq:d12new} into \eqref{eq:start_diff_eq} and assuming variation in the radial direction only, we find
\begin{equation}
w_1 - w_2 = -\frac{1}{x_1x_2}\frac{b}n\left[\frac 1p \left(\frac{d  p_1}{dr} - \rho_1  F_1' \right) + k_T\frac{d \log T}{dr} \right]
\end{equation}
where $w_i$ is the radial velocity component of $\mathbf u_i$. 
Using the ideal gas law for each species $p_i = n_i k_B T$, and writing $g'=-F_1'$, we can write
\begin{equation}
w_1 - w_2 = -\frac{b}{ n_2}\left[ \frac{d \log n_1}{dr} + \frac{m_1 g'}{k_BT}  +\left(1+\frac {k_T}{x_1}\right) \frac{d \log T}{dr} \right].\label{eq:wdiff}
\end{equation}
This equation is the same as (1) in \cite{Hunten1987}, except that our acceleration term is $g' = g + {D_0\mathbf u_0}/{Dt}$. We think that the additional ${D_0\mathbf u_0}/{Dt}$ term is unlikely to cause order of magnitude differences to the results in most cases, although we leave detailed investigation of its importance to future work.

\begin{figure}[h!]
	\begin{center}
\ifarxiv
		{\includegraphics[width=3.0in]{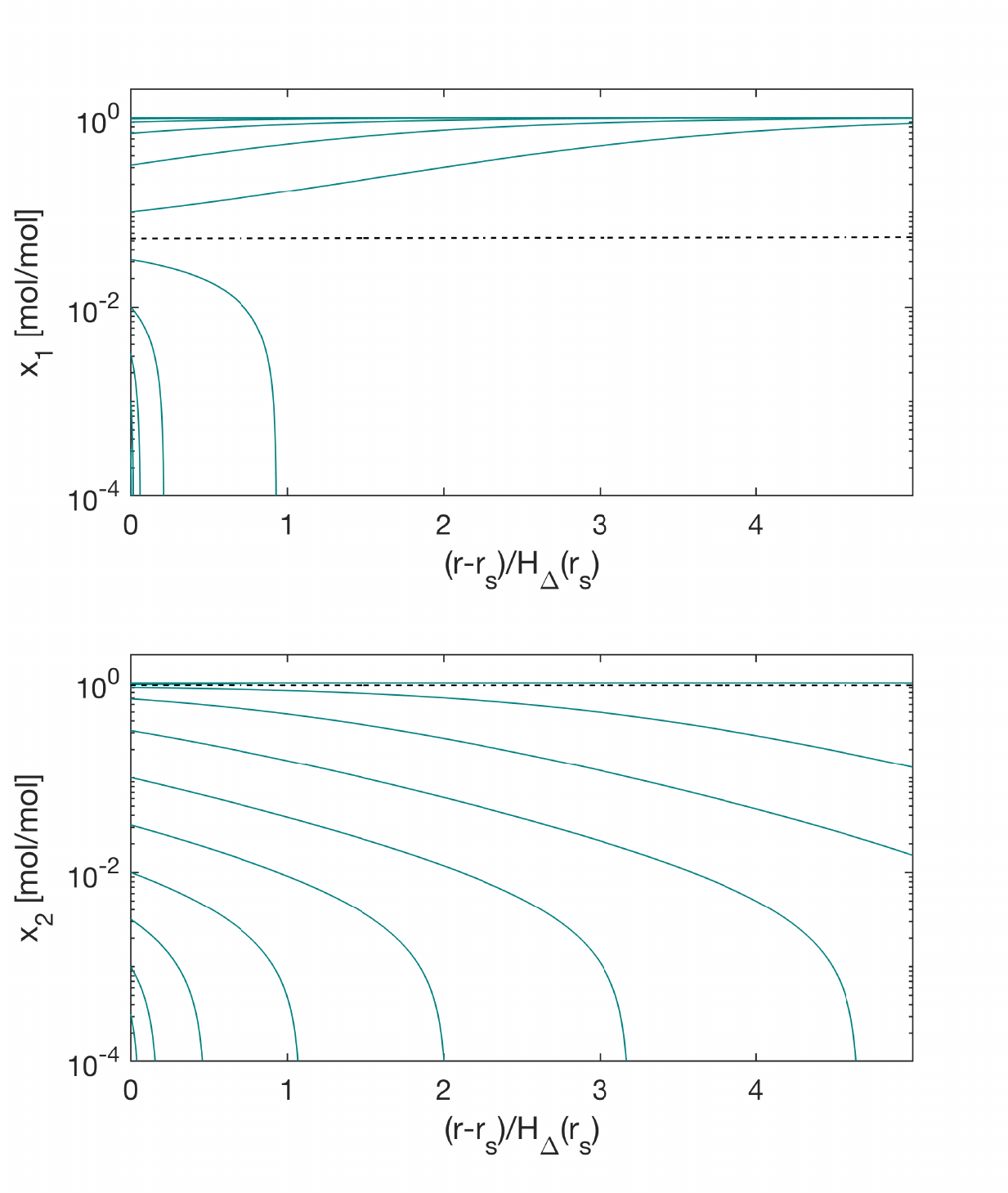}}
\else
		{\includegraphics[width=3.0in]{figures/integrate_x2_eqn_v3.pdf}}
\fi
	\end{center}
	\caption{Variation in molar concentration of a light ($x_1$; top) and heavy ($x_2$; bottom) escaping species, as determined by numerical integration of (\ref{eq:logx2_2}). Results are shown for an Earth-mass planet, $T=400$~K, molecular masses and binary diffusion coefficient appropriate for H and O (Table~\ref{tab:binary_diff}), $\Phi_1 = 1\times10^{12}$~molecules/cm$^2$/s and $\Phi_2 = 0.1 \Phi_1$. The various lines show different starting conditions for $x_1$ at $r=r_s$.}
\label{fig:integrate_x2_eqn}
\end{figure}

Next, we define molecule number flux per unit surface area as $\Phi_i = n_i w_i (r/r_s)^2$. We also drop the thermal diffusion term involving $d\log T/dr$, as previous work \citep{Zahnle1986b} has shown it is generally small. Rearranging \eqref{eq:wdiff}, we find 
\begin{equation}
\frac{d  n_1}{dr} = \frac 1 {b}\frac{r_s^2}{r^2}\left(\Phi_2 n_1 - \Phi_1n_2\right) - \frac{n_1}{H_1(r)}  \label{eq:dn1dr} 
\end{equation}
and by symmetry
\begin{equation}
\frac{d  n_2}{dr} = \frac 1 {b}\frac{r_s^2}{r^2}\left(\Phi_1 n_2 - \Phi_2n_1\right) - \frac{n_2}{H_2(r)} \label{eq:dn2dr}  
\end{equation}
where $H_i(r) = k_BT(r)/m_i g'(r)$ is the local scale height for species $i$. Summation of \eqref{eq:dn1dr} and \eqref{eq:dn2dr} yields an equation for the total number density of the escaping flow
\begin{equation}
\frac{d  n}{dr} =- \frac{n}{\overline H(r)},
\end{equation}
where $\overline H(r) = k_BT(r)/\overline m g'(r)$ and $\overline m = m_1x_1+m_2x_2$ is the local mean molar mass. Finally, by expressing $d\log x_2 / dr$ in terms of $n$ and $n_2$ and substituting \eqref{eq:dn1dr} and \eqref{eq:dn2dr} into the result, we can derive 
\begin{eqnarray}
\frac{d \log x_2}{dr} &=& \frac 1 {b}\frac{r_s^2}{r^2}\left[\Phi_1 - \Phi_2(1-x_2)/x_2\right]  \nonumber\\
                                 & & + \frac{ g'}{k_B T}(m_1-m_2)(1-x_2)\label{eq:logx2}
\end{eqnarray}
Defining $H_\Delta \equiv k_BT/(m_2-m_1)g'(r)$ and \mbox{$Y_i \equiv b^{-1}(r_s^2/r^2) H_\Delta \Phi_i $}, 
we can non-dimensionalize (\ref{eq:logx2}) as 
\begin{equation}
H_\Delta\frac{d  x_2}{d  r} = +x_2^2 +  (Y_1 + Y_2 - 1) x_2 - Y_2. \label{eq:logx2_2}
\end{equation}
Assuming that $H_\Delta$ remains finite, local solutions where $dx_2/dr = 0$ are defined by
\begin{equation}
x_2^* = \frac12(1-Y_1-Y_2 \pm \sqrt{(1-Y_1-Y_2)^2 + 4Y_2})
\label{eq:logx2_3}
\end{equation}
Physical constraints require $0<x_2( r)\le 1$ for all $ r$, which allows us to discard the  solution with the negative square root. In addition, an analytical stability analysis of (\ref{eq:logx2_2}) (not shown here) reveals that the remaining root is in fact an unstable solution. This behavior is seen clearly in Figure~\ref{fig:integrate_x2_eqn}, which shows the results of numerical integration of (\ref{eq:logx2_2}) in a representative case for a range of starting values.
All starting values of $x_2$ except for one result in either $x_2 \to 0$ or $x_1 \to 0$ at large $r$, neither of which is consistent with having finite escape fluxes $\Phi_1$ and $\Phi_2$. For isothermal atmospheres where the acceleration correction to $g$ is small, the correct starting value is simply $x_2^*$, but in general situations it can vary, as for the case shown in Figure~\ref{fig:integrate_x2_eqn}. In realistic flows, both $T$ and $g'$ should increase with altitude, although typically not by enough to change $x_2^*$ by more than a factor of a few. 

The key point of Figure~\ref{fig:integrate_x2_eqn} is that the cases where $x_2$ goes to 0 or 1 at large radii are inconsistent with the escape of both species. Only the starting values of $x_2$ that are close to $x_2^*$ keep $0<x_2(r)<1$ for all $r$. Physically, we can expect that in a temporally evolving flow, the value of $x_2$ near the base would adjust until this constraint were automatically satisfied.

\begin{figure}[h]
	\begin{center}
\ifarxiv
		{\includegraphics[width=3.0in]{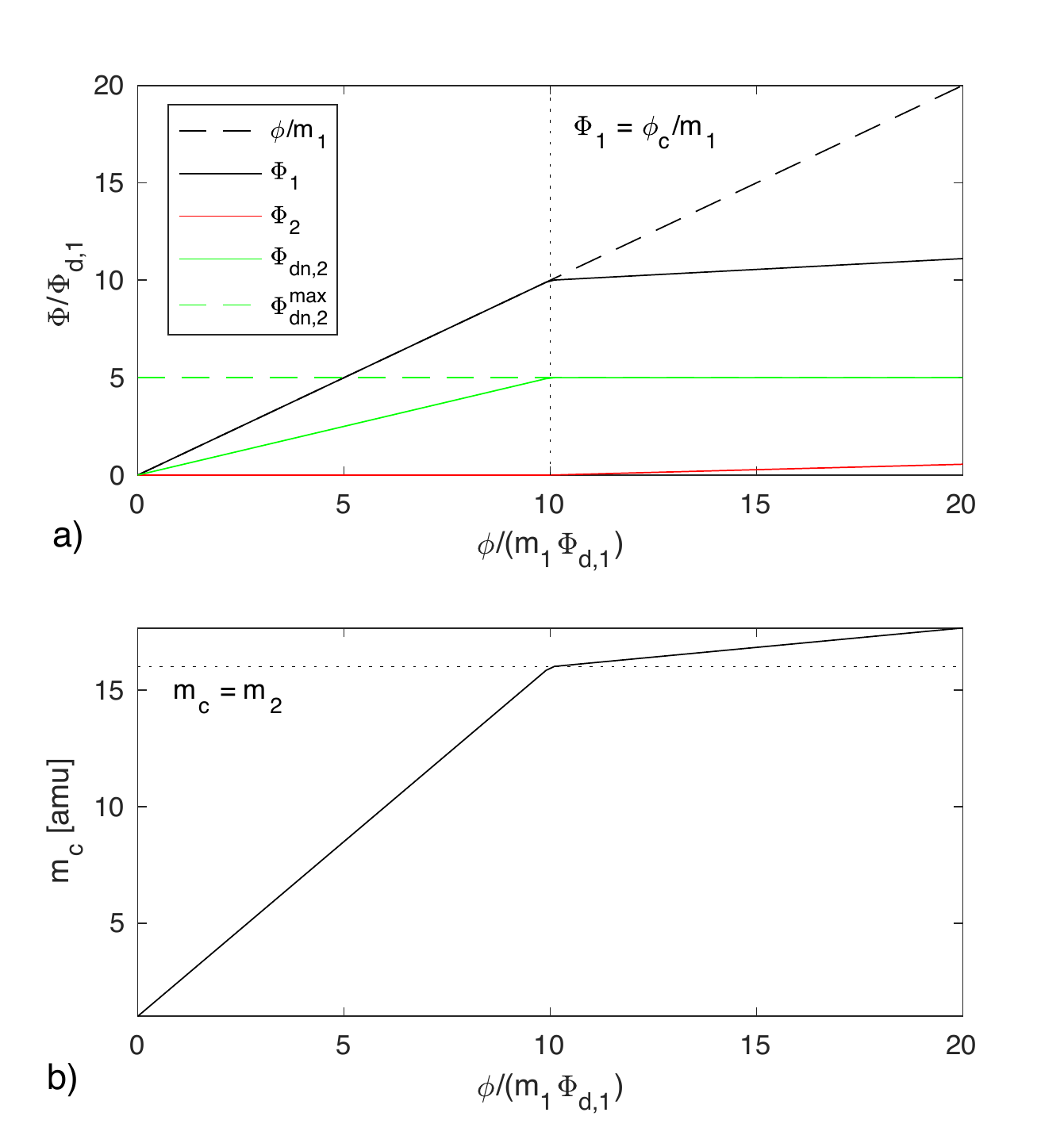}}
\else
		{\includegraphics[width=3.0in]{figures/O_drag_analytic_v2.pdf}}
\fi
	\end{center}
	\caption{a) Normalized number fluxes from \eqref{eq:num_flux_1}, \eqref{eq:num_flux_2} and \eqref{eq:num_flux_3} and b) crossover mass from \eqref{eq:crossover_m} vs. the normalized total mass flux $\phi$. Here $m_1=1$~amu, $m_2=16$~amu, $x_1=2/3$ and $x_2=1/3$. The dashed line in a) indicates that drag of species 2 (atomic O) commences when the crossover flux is reached, while b) shows  the equivalent criterion $m_c=m_2$.}
\label{fig:O_drag_analytic}
\end{figure}

By mass conservation, the total mass flux must equal the combined mass fluxes of the two species
\begin{equation}
\phi = m_1\Phi_1+ m_2\Phi_2.\label{eq:masscons}
\end{equation}
When the variations of $g'$ and $H_\Delta$ with $r$ are small, we can substitute \eqref{eq:masscons} into \eqref{eq:logx2_3} and rearrange to derive (\ref{eq:num_flux_1}) and (\ref{eq:num_flux_2}), given (\ref{eq:diff_flux}).

We regard this approach as a significant improvement over the commonly used crossover mass approach of \cite{Hunten1987}, because it starts from first principles, making all approximations clear, and allows us to express the fluxes of species 1 and 2 directly in terms of externally defined quantities. The crossover mass itself can still be used as a diagnostic, as it can be written based on the definition of $\Phi_{d,1}$ as 
\begin{equation}
m_c = m_1\left(1  + \frac{\Phi_1 }{ \Phi_{d,1} x_1}\right).\label{eq:crossover_m}
\end{equation}
More usefully, the \emph{crossover flux} $\phi_{c}$ defining the threshold for escape of the heavy species 2  is simply equivalent to the value of $\Phi_1 m_1$ when $m_c = m_2$ [see (\ref{eq:cross_flux}) in the main text]. For the water loss problem, $\phi_{c}/m_1 = 10\Phi_{d,1}$, or about $1.3\times 10^{13}$~atoms/cm$^2$/s for an Earth-mass planet. 
For energy-limited XUV-driven escape with efficiency $\epsilon = 0.15$, this requires an XUV flux of around 0.35~W/m$^2$, or around 100~times that incident on Earth today. 
Figure~\ref{fig:O_drag_analytic} shows plots of fluxes and the crossover mass vs. the species 1 reference flux normalized by its diffusion flux, demonstrating the simplicity of our approach.

Finally,  the net build-up of species 2 in the planet's atmosphere can be written in general as
\begin{equation}
\Phi_{\downarrow,2} = \Phi_1 (x_2/x_1) - \Phi_2 .      \label{eq:num_flux_prev}
\end{equation}
Hence 
\begin{eqnarray}
 \Phi_{\downarrow,2}  \approx \left\{
  \begin{array}{lr}
   (x_2/x_1) \phi / m_1 & :  \phi <\phi_c\\
   {x_2b}(H_2^{-1} - H_1^{-1})  & : \phi  \ge \phi_c.
  \end{array} \right.
      \label{eq:num_flux_3}
\end{eqnarray}
Given $x_1=2/3$, $x_2=1/3$, $m_1 = 1$~amu and $m_2 = 16$~amu, it is easily shown that $ \Phi_{\downarrow,2}$ becomes
\begin{equation}
\Phi_{\downarrow,2} \approx \frac{5b m_p g}{k_B T},\label{eq:phidownarrow}
\end{equation}
in agreement with previous work \citep{Zahnle1986,Luger2015,Tian2015b}.

\section{Analytic limit on pre-main sequence magma ocean \ce{O2} buildup}\label{apx:MO_buildup}

To ensure that we understand the results of the coupled escape-climate-interior model described in Section~\ref{sec:melty}, here we derive an analytic estimate of the planet composition for which \ce{O2} buildup ceases to occur. First, we parametrize the pre-main sequence stellar luminosity as
\begin{equation}
L(t)=  L_0 (t/t_0)^\alpha  
\end{equation}
where $L_0$ is the luminosity at time $t_0$, which we define as the start of the star's main sequence phase. Based on a least-squares fit of the \cite{Baraffe2015} data, we have found $\alpha = -0.725$, $L_0 = 8.3 \times10^{-4} L_S$ and $t_0 = 0.43$~Gy for a $0.1M_S$ red dwarf star.

Next, we assume that  \ce{O2} buildup will start to occur when the total oxidation due to preferential H escape  is greater than the total amount of FeO in the magma ocean. 
Rather than performing an integral, we simply estimate the oxidation due to escape as $Et_{RG}=4\pi r_P^2 t_{RG}\Phi_{\ce{H},UV}$, with $t_{RG}$ the time at which the runaway greenhouse state ceases. Hence the required melt fraction is
\begin{equation}
\Psi_{RG}=\frac{  4\pi r_P^2 t_{RG}\Phi_{\ce{H},UV} }{f_{\ce{Fe_2O3},max}N_{\ce{FeO},0}}.
\end{equation}
Here $f_{\ce{Fe_2O3},max}$ is the maximum level that \ce{Fe_2O3} is allowed to build up to in the magma, which we take to be 0.3 based on Fig.~\ref{fig:outgas_redox}. Noting also that $N_{\ce{FeO},0}=q_{\ce{FeO},0}M_P(1-f_c)/m_{\ce{FeO}}$, we can use the analytic expression (\ref{eq:global_melt_fraction}) for $\Psi_{RG}$ and solve for $T_s$ to get 
\begin{equation}
T_{s,RG} = T_t + \Delta T \mbox{erfinv}\left[2 \Psi_{RG} - 1\right].
\end{equation}
the minimum surface temperature required at time $t_{RG}$ to avoid \ce{H2O} buildup. 

The final step is to relate $T_{s,RG}$ to the minimum required starting \ce{H2O} inventory. To do this we first assume that the atmosphere behaves like an optically thick gray emitter \citep{Pierrehumbert2011BOOK}
\begin{equation}
T_{s,RG}=T_{e,RG}\left(\frac{p_e}{p_{s,RG}}\right)^{R/c_p},\label{eq:drythickadia}
\end{equation}
with 
\begin{equation}
T_{e,RG} \approx \left(\frac{282\mbox{ W/m}^2}{\sigma_B}\right)^{1/4},
\end{equation}
and $p_e = 0.05$~bar based on comparison with the LBL OLR data. Here $R$, $c_p$ and $\sigma_B$ are the specific gas constant, heat capacity at constant volume and Stefan Boltzmann constant, respectively.

The total mass of \ce{H2O} required is then the amount lost to space up to time $t_{RG}$, plus the amount required in the atmosphere and melt required to sustain surface temperature $T_{s,RG}$, i.e.
\begin{equation}
M_{\ce{H2O},all}  = M_{lost} + M_a + M_v.\label{eq:anal_soln_a}
\end{equation}
(\ref{eq:anal_soln_a}) can be expanded and rearranged as 
\begin{equation}
q_{\ce{H2O},0}  =   \frac{M_{lost}}{M_p} + \frac{4\pi r_p^2 p_{s,RG}} { gM_p} + (1-f_c)q_{ref}\left(\frac{p_{s,RG}}{p_{ref}}\right)^\beta.\label{eq:anal_soln_b}
\end{equation}
$p_{s,RG}$ is defined by (\ref{eq:drythickadia}), while $M_{lost}$ is taken from the model results. Despite the number of approximations that have gone into (\ref{eq:anal_soln_b}), reference to Fig.~\ref{fig:PMS_O2_buildup} shows that it reproduces the numerical results quite well.

\section{Photochemical data}

\begin{table*}[h]
\centering
  \begin{adjustbox}{max width=\textwidth}
  \begin{tabular}{crlcc}
  \hline
  \hline
\# &  Reaction &  & Rate coefficient & Reference \\
  \hline
A1 &  \ce{H2} & \ce{->C[h\nu] 2H} & From cross-section data & 1  \\
A2 &  \ce{O2} & \ce{->C[h\nu] 2O} & From cross-section and quantum yield data & 2 \\
A3 &    & \ce{->C[h\nu] O + O(^1D)} & From cross-section and quantum yield data & 3 \\
{A4} &  \ce{H2O} & \ce{->C[h\nu] OH + H} & From cross-section and quantum yield data & 4 \\
{A5} &    & \ce{->C[h\nu] H2 + O(^1D)} & From cross-section and quantum yield data & 4 \\
{A6} &   & \ce{->C[h\nu] 2H + O} & From cross-section and quantum yield data & 4 \\
{A7} &   \ce{OH} & \ce{->C[h\nu] O + H} & From cross-section and quantum yield data &  4 \\
\hline
B1 &\ce{H + O3}&\ce{-> OH + O2}& $1.4\times 10^{-10}\E^{-470/T}$  & 5   \\
B2 &\ce{H + HO2}&\ce{-> 2OH }& $7.3\times 10^{-11} $  & 5   \\
B3 &\ce{ 		}&\ce{-> O + H2O}& $2.4\times 10^{-12} $  & 5   \\
B4 &\ce{ 		}&\ce{-> H2 + O2}& $5.6\times 10^{-12} $  & 5   \\
B5 &\ce{O(^1D) + O2}&\ce{->  O + O2}& $3.2\times 10^{-11}\E^{70/T}$  & 5   \\
B6 &\ce{O(^1D) + O3}&\ce{->  2O2}& $1.2\times 10^{-10} $  & 5   \\
B7 &\ce{                    }&\ce{->  O2 + 2O }& $1.2\times 10^{-10} $  & 5   \\
B8 &\ce{O(^1D) + H2}&\ce{-> H + OH }& $1.2\times 10^{-10} $  & 5   \\
B9 &\ce{O(^1D) + H2O}&\ce{-> 2OH }& $1.63\times 10^{-10}\E^{60/T} $  & 5   \\
B10 &\ce{O + OH}&\ce{-> O2 + H}& $1.8\times 10^{-11}\E^{180/T}$  & 5   \\
B11 &\ce{O + HO2}&\ce{-> OH + O2}& $3.0\times 10^{-11}\E^{200/T}$  & 5   \\
B12 &\ce{O + H2O2}&\ce{-> OH + HO2}& $1.4\times 10^{-12}\E^{-2000/T}$  & 5   \\
B13 &\ce{O + O3}&\ce{->  2O2}& $8\times 10^{-12}\E^{-2060/T} $ & 5   \\
B14 &\ce{2OH}&\ce{-> H2O + O}& $1.8\times10^{-12}$  & 5   \\
B15 &\ce{OH + O3}&\ce{-> HO2 + O2}& $1.7\times 10^{-12}\E^{-940/T}$  & 5   \\
B16 &\ce{OH + H2}&\ce{-> H2O + H}& $7.7\times 10^{-12}\E^{-2100/T}$  & 5   \\
B17 &\ce{OH + HO2}&\ce{-> H2O + O2}& $4.8\times 10^{-11}\E^{250/T}$  & 5   \\
B18 &\ce{OH + H2O2}&\ce{-> HO2 + H2O}& $1.7\times 10^{-12}$  & 5   \\
B19 &\ce{HO2 + O3}&\ce{-> OH + 2O2}& $1.0\times 10^{-14}\E^{-500/T}$  & 5   \\
B20 &\ce{HO2 + HO2}&\ce{-> H2O2 + O2}& $2.3\times 10^{-13}\E^{600/T}$  & 5   \\
  \hline
C1 &\ce{O + O}&\ce{->C[M] O2}& $k = 1.1\times10^{-27} T^{-2}$  & 6 \\
C2 &\ce{H + H}&\ce{->C[M] H2}& $k = 1.8\times10^{-30} T^{-1}$  & 5 \\
C3 &\ce{H + O2}&\ce{->C[M] HO2}& $k_0 = 1.3\times10^{-27} T^{-1.6}$, $k_{\infty} = 7.5\times10^{-11}$  & 6 \\
C4 &\ce{OH + OH}&\ce{->C[M] H2O2}& $k_0 = 1.7\times10^{-28} T^{-0.8}$, $k_{\infty} = 1.5\times10^{-11}$  & 6 \\
C5 &\ce{HO2 + HO2}&\ce{->C[M] H2O2 + O2}& $k = 1.2\times10^{-31}$  & 6 \\
C6 &\ce{O + O2}&\ce{->C[M] O3}& $k = 2.989\times10^{-28} T^{-2.3}$  & 5 \\
  \hline
  \hline
\end{tabular}
\end{adjustbox}
\caption{Reactions used in the photochemical calculations: (A) are photolysis reactions, (B) are 2-body (units of cm$^3$/molecule/s) and (C) are 3-body (units of cm$^6$/molecule$^2$/s). Photodissociation cross-sections and quantum yields are taken from the same sources as in \cite{Venot2012}, with the exception of \ce{O2}. References are 1: \cite{Samson1994,Chan1992,Olney1997}, 2: \cite{Sander2009}, 3: \cite{Brion1979,Yoshino1992,Chan1993c,Fally2000}, 4: \cite{Chan1993,Mota2005,Huebner1992}, 5: \cite{Linstrom2010}, 6: \cite{Yung1999}.}
\end{table*}

\end{appendices}

\bibliographystyle{plainnat}



\end{document}